\documentclass[12pt]{article}
\usepackage{graphicx} % Required for inserting images
\usepackage{natbib}
\usepackage{subfigure}
\usepackage{caption}
\usepackage{float}
\usepackage{graphicx}
\usepackage{hyperref}
\usepackage{url}
\usepackage{amssymb}
\usepackage{amsmath}
\usepackage{adjustbox}
\usepackage{booktabs}
\usepackage{graphicx}
\usepackage{svg}
\usepackage{chngcntr}
\usepackage{anysize}
\marginsize{2.0 cm}{2.0 cm}{2.5 cm}{2.5 cm}

\counterwithout{table}{section}

\begin{document}

\title{Go Green Without the Mafia! Dissolution of Infiltrated City Councils and Environmental Policy\thanks{We thank seminar participants at Urbino (SIE, Annual Meeting), Rome (IAERE, Annual Conference), Enna (CiMET, XXI Workshop), for useful comments. This study was funded by the European Union - NextGenerationEU, Mission 4, Component 2, in the framework of the GRINS (Growing Resilient, INclusive and Sustainable project. GRINS PE00000018 - CUP B73C22001260006). The views and opinions expressed are solely those of the authors and do not necessarily reflect those of the European Union, nor can the European Union be held responsible for them. Usual disclaimer applies.}}

\author{Andrea Mario Lavezzi\thanks{Department of Law, University of Palermo, Piazza Bologni 8, 90134, Palermo, Italy. Email: mario.lavezzi@unipa.it}\\\small University of Palermo \and Marco Quatrosi\thanks{Department of Law, University of Palermo, Piazza Bologni 8, 90134, Palermo, Italy. Email: marco.quatrosi@unipa.it}\\\small University of Palermo 
}
\maketitle

\date{}
\vspace{-1cm}
\begin{abstract}
In this article, we study the effects of   organized crime infiltration in city councils on environmental policies implemented in Italy at the municipal level. To this purpose, we exploit the exogenous shock of the removal of a city council infiltrated by the mafia and its substitution with an external Commission, allowed in Italy by the law 164/1991. Our results suggest that after dissolution, environmental policies improve in several dimensions: the capital expenditure for sustainable development and the environment increases; the current expenditure on integrated water system increases; the percentage of sorted waste increases because, as we show, public expenditure is reallocated toward sorted waste at the expenses of unsorted waste. These results are robust to different specifications of the control group. In addition, we find significant spillover effects: the dissolution of infiltrated city councils implies an improvement in environmental policies in adjacent municipalities. Our results have a straightforward policy implication, the need to combat organized crime as a way to improve the environmental conditions of the territories plagued by its pervasive presence.
\end{abstract}


\section{Introduction}
Organized crime around the world poses a serious threat to economic development. Recent studies such as \citet{pinotti2015causes}, for example, identify a negative effect of organized crime on economic development in a cross-section of countries. Italy represents a peculiar case as a country where powerful and long-standing criminal organizations, such as \textit{Cosa Nostra}, the Camorra or the N'Drangheta, originated in the distant past in, respectively, Sicily, Campania or Calabria, are still able to exert a negative impact on political and economic outcomes. \citet{pinotti2015economic}, in particular, documents a significant cumulative loss of GDP in recent years in the Italian region of Apulia, amounting to approximately 16\% of its GDP, while \citet{alesina2019organized} show that Italian criminal organizations can influence the political selection of candidates and the subsequent political decisions through violence and intimidation.

The negative impact of organized crime, however, extends beyond the economic and the political dimension as criminal organizations have also been recognized as key players in the committing of environmental crimes, such as illegal waste disposal. \citet{massari2004dirty}, in particular, show that criminal organizations exploit weaknesses in regulatory frameworks to engage in environmentally harmful practices (see also \citealp{germani2018eco}). As a consequence, inefficient public environmental policies can inadvertently create a favorable environment for organized crime.

In this article, we focus on the effect that criminal organizations exert on local environmental policies in Italy, in particular through the infiltration of local municipal councils. Specifically, we conjecture that a mafia-infiltrated city council may be reluctant to implement efficient environmental policies, as this allows organized crime to profit from these inefficiencies.\footnote{In this article we use the terms ``mafia" and ``organized crime" as synonyms.} Furthermore, infiltrated councils may in general be less efficient and lack the political will or resources to enforce stricter regulations, further strengthening a vicious circle of mafia infiltration and environmental degradation.

%The interplay between organized crime and environmental policy highlights a broader challenge: the need for effective governance and law enforcement to combat these illicit activities. Strengthening institutions, enhancing transparency, and fostering community engagement are essential steps toward mitigating the socioeconomic impact of organized crime and promoting sustainable development in affected regions.

In particular, we exploit the dissolution of local councils and the appointment of an external Commission, allowed in Italy by Law 164/1991, to identify a causal link between the infiltration of local governments by organized crime and environmental policies. Our hypothesis, as sketched above, is that an infiltrated local council implements inefficient environmental policies so that the removal of the infiltrated council and its substitution with an external Commission appointed by the State is expected to improve the efficiency of the environmental policies implemented in that municipality. 

Our main results are the following: after dissolution, the environmental policies implemented by the affected municipalities improve in different dimensions. Specifically, the capital expenditure for sustainable development and the environment increases; the current expenditure on integrated water services increases; the percentage of sorted waste increases. The latter result appears to be a consequence of the reallocation of the public expenditure towards sorted waste at the expenses of unsorted waste. These results are robust to different specifications of the control group. In addition, we identify significant spillover effects: the dissolution of infiltrated city councils implies an improvement along the same dimensions in adjacent municipalities. 

%that organized crime is pervasive, mainly in areas related to waste management and sustainable waste treatments. Not only does the mafia affect the effectiveness of more sustainable waste treatment policies at the local level, but it also has repercussions on the efficiency and, ultimately, on the overall sustainability of such initiatives for municipalities. In other areas related to the environmental management of municipalities, such as water management and green areas, the redistribution of competencies among administrative levels after the introduction of the Environmental Code might have diminished the influence of organized crime, at least at the municipality level. More studies should be conducted to assess the possible influence of mafia on environmental management at higher administrative levels.

The remainder of the paper is organized as follows. In Section \ref{secLiterature} we clarify the contribution of this article with respect to the existing literature; in Section \ref{secInstitBack} we describe the institutional background: specifically, in Section \ref{secEnvPolicies} we describe the main features of environmental policies in Italy, while in Section \ref{secLaw164} we describe the law 164/1991 on the dissolution of infiltrated local councils. In Section \ref{secDataMethod} we describe the data and the empirical methodology, while in Section \ref{secResults} we present the results. Section \ref{secConclusions} contains some concluding remarks.

\section{Literature Review\label{secLiterature}}
Our contribution speaks to two main strands of literature: the one on organized crime and the environment, and the one on the social, political and economic effect of the dissolution of the city councils infiltrated by organized crime.

%\textbf{(qui vanno indicati brevemente i findings di questi articoli su organized crime and environment)}

The literature pointed out that organized crime has significant detrimental effects on environmental integrity and, in general, on sustainable development \citep{reitano2018organized}. \citet{lynch2020green} examines the intersection of organized crime and green criminology, providing a comprehensive framework for analyzing environmental harm facilitated by criminal groups. This work highlights how the lack of a universally accepted definition of ``green crime" hindered academic contributions to this issue \citep{lynch2020green}. Moreover, the diverse contexts in which green crimes occur, such as ecosystems, climate change, air and water pollution, waste management, and animal trafficking, have not been fully integrated into the criminology literature \citep{lynch2020green}.

Criminals exploit the lack of international consensus and the different approaches taken by countries to combat environmental crimes \citep{unep2016rise}. An example of this is related to the legislative framework on end-of-waste. The lack of clear norms identifying when materials can be defined as waste favored illegal waste trafficking \citep{sergi2016earth}. In this context, corruption further exacerbates environmental harm by facilitating illegal acts against the environment \citep{thompson2023convergence}. In addition, \citet{kangaspunta2009eco} discuss the socioeconomic impacts of eco-crimes, emphasizing the globalized nature of environmental crimes. Transnational environmental crimes often involve the trafficking of plants, animals, and hazardous substances across national borders, with organized crime playing an important role, including the corruption of public institutions through bribery and other means \citep{kangaspunta2009eco}.
Illicit activities that harm the environment can take an extractive form, as the trafficking in natural resources \citep{rege2017organization,zabyelina2020new,brombacher2021introduction}. Alternatively, nonextractive illicit activities, including the cultivation of drugs, also result in environmental degradation \citep{devine2021narco,horky2021methamphetamine,pardal2021synthetic}. 

In Italy, organized crime has significant impacts on environmental sustainability, particularly through illegal waste disposal and the related negative impact on environmental pollution. Between 2022 and 2023, environmental crimes in Italy increased by 15.6\% \citep{legambiente2024ecomafia}. Illegal activities within the waste cycle constitute the most prevalent environmental crime, with 9,309 offenses reported in 2023, representing an increase of 66.1\% from the previous year \citep{legambiente2024ecomafia}. \citet{sergi2016earth} explore the emergence of the ``eco-mafia," linking criminal organizations to hazardous waste trafficking and illegal landfills in the Calabria region. Beyond waste management, organized crime has also infiltrated the renewable energy sector \citep{sergi2016earth}.

The Italian region of Calabria gained national and international media attention in the early 2000s when a shipwreck containing radioactive material was discovered off its coast. A mafia informant alleged that this was the first of many such ships sunk in the regional waters \citep{BBC2009}. Although no investigation has definitively linked the shipwreck to the 'Ndrangheta, the powerful criminal organization that originated in Calabria \citep{sergi2016earth}, evidence indicates that the 'Ndrangheta and other criminal groups, including the Neapolitan Camorra, use their territories for the illegal disposal of waste \citep{sergi2016earth}. In Campania, illegal waste disposal practices have also been observed, particularly in the area known as the \textit{Terra dei Fuochi} (Land of Fires) \citep{saviano2006gomorra}. The first documented route of illegal waste trafficking in Italy was discovered in the early 1990s, in a period when such activities were not yet criminalized under Italian law \citep{massari2004dirty} (see \citealp{massari2004dirty} for an analysis of waste trafficking networks in Southern Italy, highlighting their scale and environmental impacts).

In some cases mafia-type organizations, particularly in Southern Italy, control waste collection, transportation, and legal dumps, as well as companies involved in waste management \citep{massari2004dirty}. The handling of hazardous waste, in particular, has become one of the most lucrative sectors for criminal organizations within the waste management industry. \citet{germani2018eco} examine regional variations in waste-related organized crime and their socio-environmental consequences. Their findings reveal a regional dualism in waste trafficking, with notable differences between Northern and Southern Italy. In the North-center, this kind of crimes are positively related to education whereas in the South there exists a negative relationship. Specifically, in the North-Center, waste trafficking attracts more educated perpetrators as, unlike other forms of trafficking, operators must know the market, the complex legislation and its weaknesses \citep{germani2018eco}. In addition, in the South of Italy waste trafficking is positively related to the endowments of waste management plants. Whereas in the North, waste management plants are typically well-functioning and ensure an integrated waste cycle, which implies less incentives for illegal trafficking, in the Southern regions such plants are less efficient, as their development appears oriented towards personal interests and illegal practices \citep{germani2018eco}. Furthermore, enforcement activities appear to have limited deterrence effects on waste trafficking in most Italian regions. Only high levels of enforcement are negatively correlated with waste trafficking \citep{germani2018eco}. Lastly, \citet{di2023organized} analyze the impact on waste management costs of exposure to environmental crimes at the provincial level. \citet{di2023organized} find that waste management costs tend to be higher in provinces with a higher incidence of environmental crime.

Summing up, the literature identified a tight connection between organized crime and the environment. Organized crime, in particular, tries to exploit profit opportunities coming from different environmental-related activities, in particular waste management: from the illegal disposal of hazardous substances to the management of dumps and waste transport. An inefficient institutional framework facilitates the proliferation of illegal activities, which can increase environmental degradation.

A second strand of literature, relevant for our analysis, examined the consequences of the dissolution of city councils infiltrated by organized crime in Italy. In this literature, the dissolution is treated as an exogenous shock (see Section \ref{secLaw164} for details of the legislative framework) that hits a municipality and reduces the power of the criminal organization in the municipal territory.

Some contributions focus on the consequences for economic development, including the allocation of public expenditure.
\citet{acconcia2014mafia} is the seminal contribution in this strand of research, although they consider the dissolution of city councils as an instrument for the level of public expenditure in a study of fiscal multipliers. \citet{acconcia2014mafia} show that, after the dissolution of the city council, public expenditures diminish greatly. \citet{acconcia2014mafia} conjecture that this happens as wasteful expenditures that were probably addressed to favor the interests of the local mafia groups are cut by the Commission appointed by the State, in order to restore efficiency.

\citet{di2022organized} provide evidence supporting this conjecture. Specifically, \citet{di2022organized} examined the composition of public spending in infiltrated city councils, revealing a significant diversion of public funds towards sectors with high criminal presence such as Construction and Waste disposal. In this context, \citet{galletta2017law} shows that the dissolution of city council also has spillover effects, as public investments are also reduced in neighboring municipalities. Our approach, however, allows for a reassessment of the results \citet{di2022organized} as we distinguish the public expenditure on the disposal of sorted and unsorted waste, as the latter is likely to be the one in which organized crime is interested.

\citet{fenizia2024organized} show that dissolution of infiltrated municipalities positively affects local economic growth. \citet{fenizia2024organized} argue in particular that removing infiltrated councils restores investor confidence and redirects resources to productive uses. Finally, \citet{buonanno2024all} consider the effects of the dissolution of infiltrated city councils on social capital and find that dissolution enhances community cohesion and trust, which are critical for long-term resilience against crime and are crucial factors for economic development (see, e.g. \citealp{ashraf2024economic}, for a discussion).
Other works considered the effects of the dissolution of city councils on the political process. For example, \citet{daniele2015organised} focus on the human capital of elected politicians, assuming that it is a proxy of their quality, and find that the education level of politicians increases after the dissolution. The insight here is that the human capital of elected politicians in infiltrated municipalities is maintained artificially low by the criminal organizations, for example, by manipulating the electorate and by vote buying \citep{baraldi2022self}, as criminal groups can benefit from inefficient institutions run by politicians with little human capital. Finally, other works pointed out that the dissolution of infiltrated city councils reduces some types of crime \citep{cingano2020law} and affects the public procurement process \citep{ravenda2020effects}, also by spillovers to neighboring municipalities \citep{tulli2024sweeping}.

Few studies considered municipal council dissolution to assess the potential effect of mafia infiltration on environmental performance at the municipal level. \citet{d2015waste} find that the share of sorted waste at the provincial level is negatively associated with indicators of organized crime in the nearby areas. Although \citet{d2015waste} use the information on the municipalities dissolved for Mafia infiltration, they do not exploit it as a quasi-experimental source of variation as in the present study.

Our work is close in spirit to the recent work by \citet{baraldi2024fighting}, who analyze council dissolution in the Southern regions of Campania, Calabria, Apulia and Sicily for the period 2010-2019, and exploit the dissolution as a quasi-experimental source of variation, as in this article. \citet{baraldi2024fighting} show that dissolved municipalities increase their share of recycled waste after the removal of the infiltrated council. Our work expands the analysis of \citet{baraldi2024fighting} in three directions. On the one hand, we consider a longer period of analysis and a larger number of dissolved municipalities, as we focus on the  whole Italian territory, as in \citet{cingano2020law} and \citet{tulli2024sweeping}. This allows to consider a higher number of cases of dissolution of municipal councils. Second, we analyze a larger set of proxies for the environmental efficiency of the local councils, beyond the share of sorted waste. Third, we estimate spillover effects of the dissolution of infiltrated city councils on neighbor municipalities, as in \citet{galletta2017law} and \citet{tulli2024sweeping}. In fact, as we discuss in Section \ref{secEnvPolicies}, the organization of the environmental activities of local administrations in Italy often implies the interaction of local administrative bodies with neighbors, so the dissolution of a city council can be expected to have spillover effects.

%feeds the literature on the effects of mafia presence on the environmental performances of all Italian municipalities from 2010 to  2022 via adopting a quasi-experimental approach to a set of environmental-related outcome variables. Furthermore, our methodological approach allows us to look at the effect of the treatment over time, providing a dynamic perspective on the influence of the mafia in the municipalities.

\section{Institutional background\label{secInstitBack}}
In this section we describe the legal-institutional framework relevant for our analysis. Namely, the Italian environmental legislation, that defines the competencies at the three administrative levels prevailing in Italy, national, provincial, and municipal, and the legislation on the dissolution of the municipal councils infiltrated by organized crime.

%\textbf{(io parlerei di Italian Environmental Legislation. Non credo che il nostro focus sia solo sulle aree verdi ma sulla politica ambientale in senso lato. Quella delle aree verdi e' un aspetto, come quello dello smaltimento dei rifiuti. A noi interessa chiarire cosa fanno o dovrebbero fare i comuni per l'ambiente per la legge italiana. Questo va  contestualizzato nella legislazione ambientale italiana e su come stabilisce le responsabilita' e compiti a livello nazionale, provinciale (le province contano ancora in questo contesto?), e comunale. Vedi il cappello introduttivo che ho messo a questa sezione. Quindi, idealmente, servirebbe una breve descrizione di questi livelli, e poi un approfondimento su cosa dovrebbero fare i comuni, ad es. occuparsi di aree verdi, smaltimento rifiuti ecc.)}

\subsection{The Italian Environmental Legislation\label{secEnvPolicies}}
The primary legislative framework for environmental management in Italy is the Legislative Decree 152/2006, commonly referred to as \textit{Codice dell'Ambiente} (Environmental Code). This decree defines the roles and responsibilities of administrative authorities in promoting environmental protection and optimizing the use of natural resources (Art. 1). It incorporates key European Union directives, including Directive 2001/42/EC and Directive 2014/52/EC on the environmental impact assessment of public and private projects, as well as Directive 2008/1/EC on pollution prevention and control. Regions are granted significant legislative authority in areas related to environmental protection within the boundaries of national legislation, as stipulated by Article 117 of the Italian Constitution. For example, regions are responsible for drafting plans for waste management, air and water quality, and the management of protected green areas. In addition, regions issue environmental authorizations (AIA) and carry out environmental impact evaluations (VIA) for various projects and economic activities.

Regarding the provinces, following the enactment of Law 56/2014, their competences in environmental management were significantly reduced. However, provinces continue to play a key role in addressing hydrogeological risks, monitoring the flow of hazardous waste, and promoting sustainability awareness campaigns among citizens.

% \footnote{Law 56/2014 technically abolished the provinces, although a referendum to remove them from the Consitution failed. Currently, therefore, they still exist as administrative units.}

%\url{https://www.arera.it/atlante-per-il-consumatore/rifiuti/il-servizio-di-gestione-rifiuti/cose-e-come-e-organizzato/cosa-sono-gli-enti-di-governo-degli-ambiti-territoriali-ottimali-egato}

The \textit{Codice dell'Ambiente} also introduced specialized local administrative bodies for environmental management known as \textit{Enti di Governo dell’Ambito Territoriale Ottimale} (EGATO).\footnote{EGATO substituted the \textit{Autorità d'Ambito} instituted by the Environmental Code and dismantled by law 42/2010. The process of implementing the EGATO on the whole national territory is still incomplete (see \url{https://www.arera.it/}).} According to Article 23 of Legislative Decree 22/1997 (the so-called \textit{Decreto Ronchi}), regions are required to define geographical areas for the organization of public services, including waste management. These areas are referred to as \textit{Ambiti Territoriali Ottimali} (ATO, Optimal Territorial Areas). The boundaries of ATOs can vary within regions on the basis of administrative and logistic considerations. ATO is assigned to an EGATO that manages environmental services in that area. In the context of waste management regions have the authority to delegate the management of ATO to EGATO entities, which oversee the whole waste management process, including waste collection, treatment, and disposal. For water management, EGATO are entitled to provide an efficient organization of the Integrated Water Service.

Municipalities serve as the final link between these administrative levels and citizens. Their responsibilities include providing waste collection and transportation services, monitoring and maintaining the local water network, often in collaboration with EGATOs, and designing and maintaining urban green spaces as part of urban planning. By bridging higher administrative levels and local communities, municipalities play a critical role in ensuring the effective delivery of environmental services and fostering environmental stewardship at the grassroots level. In what follows, we describe two areas of environmental management in which local administrations have competences, water services and waste management, and show that substantial heterogeneity exists among territorial units in Italy.

The Italian legislative framework for water management has been significantly influenced by the European Directive 2000/60/CE, which provides the main guidelines for an integrated water service aimed at preserving the water resources of each Member State. The Directive stipulates that each river basin is assigned to a district responsible for overseeing its status (Art. 3). These districts are tasked with preserving the river basins under their management and monitoring human interventions to mitigate pollution risks. 
%\textbf{(watersheds sarebbero le dighe? Sarebbe meglio dams)} 
Articles 63-64 of \textit{Codice dell'Ambiente} institutes the  \textit{Autorità di Bacino} (Basin Autority), which oversees groups of river basins in specific geographical areas.

Articles 147-158 of Legislative Decree 152/2006 deals with the Integrated Water Service. Water Services are organized according to the ATO structure, and each territorial area is assigned to an EGATO, which oversees the Integrated Water Services of that area (art. 147). Article 153 of DL 152/2006 requires the transfer of ownership of water infrastructure from local authorities to the EGATO. However, within this framework, local authorities, such as municipalities, can allocate expenses to carry out necessary works to upgrade the water service in areas already urbanized, provided these efforts comply with the EGATO’s plans (Art. 147 of DL 152/2006).

\begin{figure}[H]
    \centering
    \includegraphics[width=0.6\linewidth]{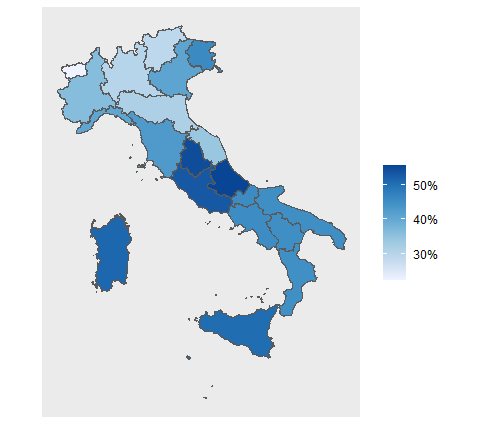}
    \caption{Water leakages in Italian Regions in 2018.}
    \label{fig:leakages}
\end{figure}

To gauge the different level of efficiency in water system management that characterizes the Italian territory, Figure \ref{fig:leakages} presents an overview of the extent of water leakage in Italy's water system in 2018 (data are from the National Census of Water for civilian use maintained by ISTAT, the Italian National Statistical Institute). Figure \ref{fig:leakages} highlights that substantial heterogeneity exists acrosss Italian regions. Specifically, the map in Figure \ref{fig:leakages} shows that the areas most affected by water leakage are concentrated in Central and Southern Italy, with some regions exceeding 50\% of leakages in their local water infrastructures. 

Waste management policy has been designed to promote a drastic reduction of waste flows as a key environmental management objective (COM(2019/640) final). To achieve those objectives, set at the European Level, the National Recovery and Resilience Plan (NRRP) envisioned the National Strategy for the Circular Economy. In particular, the legislative decree 152/2020 introduced the \textit{Piano nazionale per la gestione dei rifiuti} (National Plan for Waste Management). This plan aims to increase the share of sorted waste and improve the collection of specific waste types (e.g., RAEE, plastic, etc.). 
%\textbf{(controllare questo paragrafo, che il senso sia corretto)}

The National Plan for Waste Management provides regions and provinces with the criteria and strategies to design their plans. Each regional authority is required to issue a Regional Program for Waste Management that details objectives in line with the National Circular Economy Strategies. In particular, within the Operational Agreements of the NRRP, the National Plan of Waste Management sets specific objectives in order to promote sorted waste collection.\footnote{For details, see: \texttt{https://eur-lex.europa.eu/legal-content/EN/TXT/PDF/?uri=CONSIL:ST\_10160\_2021\_ADD\_1\_REV\_2}.}

%\footnote{Other objectives include: iii)  within December 31, 2023, the number of illegal landfills according NIF 2003/2007 procedure should have been reduced from 33 to 6; within December 31 2023 the number of illegal landfills according to NIF 2011/2015 procedure should have been reduced from 34 to 14}

 %: i) within December 31, 2023, the difference between the national average of sorted waste collected and the worst-performing Regions should have reduced by 20\% with respect to the value in 2019 (22,8\%); ii) within December 31, 2024, the variation of the average flow of sorted waste collected by the best-performing regions and the worst-performing Regions should have been reduced by 20\% with respect to the value in 2019 (27,6\%).
 
The \textit{Codice dell'Ambiente} details the competencies of each administrative authority throughout the waste management process. In particular, the law assigns full regulatory authority over waste management, including the collection, separation, and treatment of urban and hazardous waste, to the regions (Art. 196(b)). Regions are also empowered to approve projects for new waste treatment facilities, including those for hazardous waste, and to modify existing facilities (Art. 196(e)). However, Art. 195(a) assigns the State the responsibility of identifying facilities for waste treatment and recovery that are essential for achieving national objectives.

Furthermore, regions are responsible for issuing authorizations for the treatment and recycling of waste (Art. 196(e)), while the provinces have the competences of controlling and monitoring the flow of urban and hazardous waste (Art. 197). In this context, EGATO should minimize waste exchange and ensure self-sufficiency in the collection, transportation, and treatment of waste. Finally, municipalities are responsible for waste collection, ensuring the separation of different waste streams (urban and hazardous) and optimizing waste collection and transportation (Arts. 198(c-d)).

This overview clearly shows that national, regional, and municipal administrative levels are involved in some key environemntal policies. Municipalities, the main focus of this article, in particular contribute to waste management, especially to waste collection and transportation, through local regulations (art. 198 subsection 2, Legislative Decree 152/2006). Municipalities can also manage waste collection and treatment through consortia as envisioned by Art. 31 of the Legislative Decree 267/2000 and subsection 7 of Art. 200 of the Environmental Code. Waste management consortia are forms of association between local authorities at the provincial or sub-provincial level, created to ensure a more efficient and integrated management of waste collection and treatment services. In some cases, consortia have been absorbed by EGATO; in others, they exist with mainly operational competences. As regions have the right to regulate waste management at the local level (Art. 196 of the Environmental Code), the competences of each consortium may vary region by region.

\begin{figure}[H]
    \centering
    \includegraphics[width=1\linewidth]{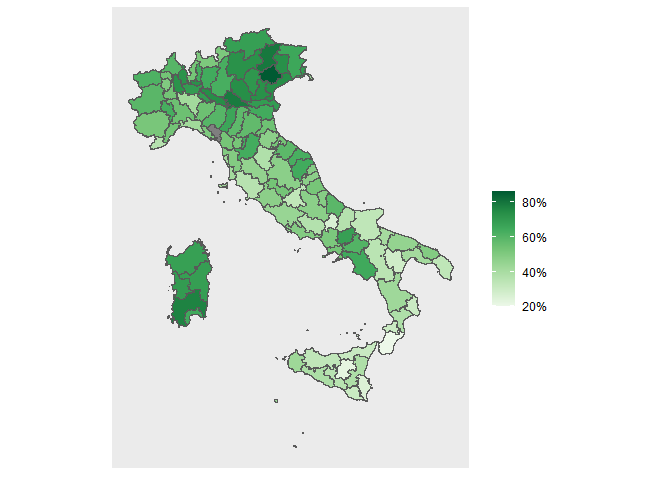}
    \caption{Percentage of sorted waste in Italian Provinces}
    \label{fig:map_sortedw}
\end{figure}

In terms of waste management, as in the case of water system management, despite the fact that key objectives are set at the National level, the Italian territory displays a great deal of heterogeneity, as Figure \ref{fig:map_sortedw} highlights with respect to the percentage of sorted waste.

%\begin{figure}
 %   \centering
  %  \includegraphics[width=1\linewidth]{map_dump.png}
   % \caption{Waste destined to Landfilling (tonnes) in the period 2010-2022. The red dot represents the Province where the facility is located }
    %\label{fig:dump}
%\end{figure}

The provinces with the poorest performance in terms of sorted waste are often located in Southern Italy (although some provinces stand out as good performers), while in Northern Italy and in Sardinia some provinces reach a percentage of sorted waste of almost 80\%. The next section describes the law allowing the dissolution of infiltrated city councils.

\subsection{The law 164/1991\label{secLaw164}}
In 1991, the Italian Parliament approved a law (D.L. 31/05/1991, n. 164) to break the links between the local public administration and organized crime.  In order to fight political corruption induced by criminal organizations in local administration, this law empowers the central government to replace infiltrated city councils with a three-person commission composed of experienced officials. The commission is endowed with full executive and legislative authority for an initial period of usually 18 months, which can be extended to a maximum of 24 months. After this period, a round of local elections is held to install a new council. 

The process involves a police investigation to identify possible links between the municipality and the criminal organizations. Once the information has been gathered, the police reports to the \textit{Prefetto}, i.e. the provincial representative of the Ministry of Internal Affairs. The \textit{Prefetto} assigns a commission to effectively find such links and report whether the municipality is liable to be dissolved for mafia infiltration. Within 45 days, the \textit{Prefetto} notifies the Ministry of Interior, which will decide if there is ground for the council's dissolution. Once the decision is taken, the President of the Republic issues the decree of council dissolution.

\begin{figure}[H]
    \centering
    \includegraphics[width=0.6\linewidth]{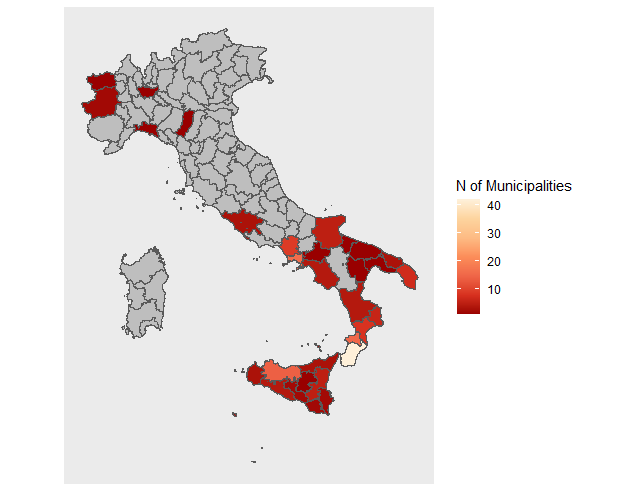}
    \caption{Number of Dissolved Municipalites for mafia inflitration in the Period 2010-2022 in Italy}
    \label{fig:dissolved}
\end{figure}

The application of the law brought to the dissolution for mafia infiltration of 169 municipal councils in the period 2010-2022. Figure\ref{fig:dissolved} shows the distribution of the affected municipalities across the national territory. 

Figure \ref{fig:dissolved} highlights that the higher number of dissolved municipalities are located in Southern Italy, although there are also episodes of dismissals due to mafia infiltration in the Central and Northern Italy signaling that to some extent the issue of organized crime infiltration is present not only in the South. It is well-known, in fact, that Italian mafias move to other Italian and foreign territories, in search of profit opportunities \citep{varese2011mafias,buonannoPazzona2014}.

The evidence on the territorial heterogeneity in the efficiency of environmental policies, both in terms of the water services and the waste management, and of the location of dissolved municipalities suggests the possible correlation between these two phenomena and allows to conjecture that organized crime might contribute reduce the level of environmental policy efficiency in particular, as we argue here, by reducing the efficiency of municipal councils in implementing efficient environmental policies.\footnote{Anecdotal evidence on the province of Agrigento, in Sicily, highlighted that the Sicilian Mafia infiltrated the local private company that was in charge of the water services, \textit{Girgenti Acque}. This rare case of privatization of water services could be considered as a response to the inefficiency of the public local services, suggesting that infiltrating a private company taking charge of a privatized environmental service, can be another channel through which organized crime benefits from inefficient environmental policies. See, e.g.: \url{https://www.agrigentonotizie.it/cronaca/girgenti-acque-tar-interdittiva-commissariamento-legittimi-no-risarcimento-danni.html}} In the next section we describe the dataset and the empirical methodology we utilize to corroborate this hypothesis.

\section{Data and Empirical Methodology\label{secDataMethod}}

In this Section we describe the data and the empirical methodology implemented in this article.

\subsection{Data\label{secData}}
Our dataset covers the period 2010-2022, and contains two types of data: i) budget data on environmental expenditures at municipal level; ii) measures of environmental performance and of environmental activities' costs incurred by the municipality, with a specific focus on costs for waste management, sorted and unsorted. We consider both types of data relevant to assess the efficiency of the implementation of environmental policies at municipal level. The first set of data allows to evaluate one aspect of the allocation of the municipal budget to environmental activities, i.e. its subdivision between investments in environmental protection in general and other activities such as water services and green areas. The second set of data refers more directly to waste management, an area that we know is very sensitive to the interests of organized crime.\footnote{In addition to national, European, or regional transfers, the main sources of financing municipal current expenditures in environmental services are local taxes. For example, waste management at the municipal level is mostly funded through the TARI (\textit{Tassa sui Rifiuti}), whereas green urban planning draws resources from the real estate IMU tax (\textit{Imposta Municipale Unica}). Finally, water management is financed by a specific tax for the integrated water service (\textit{Servizio Idrico Integrato}), which is used for the maintenance of the local water system. 
%\textbf{(punto sui finanziamenti interessante. Per ora in nota, se non troviamo una collocazione migliore.)} 
}

Specifically, we extracted current and capital expenditures for selected areas of activity of Municipalities from the OpenBDAP repository of the Italian Ministry of Internal Affairs and \textit{Openbilanci} (\url{https://openbilanci.it/}) \textit{OpenBilanci} is an initiative funded by the Lazio Region that provides information on the budget of Italian municipalities from 2010 to 2022. The Italian public sector budgeting is classified into Missions and Programs. Missions define the overarching policy objectives pursued by the local authority, such as environmental protection, local development, or urban planning, and are consistent with national classifications. Within each mission, programs identify coherent sets of activities carried out by the municipality to achieve those objectives. Programs also constitute the legal units of appropriation and administrative accountability in the municipal budget.

The data retrieved from these sources are: i) (per capita) capital expenditures for sustainable development and protection of the territory and the environment (Mission); ii) (per capita) current municipal expenditure for the integrated water service (Program); iii) (per capita) current municipal expenditures on green urban areas (Program). Capital expenditures are aimed at financing long-term investments and the creation or enhancement of public assets, whereas current expenditures are intended to cover the day-to-day functioning of public services. For example, under Mission 1 capital expenditures (i) for sustainable development and protection of the territory and the environment can include the construction of waste treatment plants, the development of flood prevention infrastructure (such as retention basins), or the installation of photovoltaic systems in municipal buildings, investments that generate value over time and contribute to environmental resilience. In contrast, current expenditures within the same mission, particularly for (ii) integrated water service and (iii) urban green areas, include the payment of utility bills, routine maintenance of water supply networks, seasonal tree pruning, lawn mowing, and irrigation services. This functional and economic classification of expenditures is crucial not only for the internal efficiency of municipal management, but also for evaluating the long-term sustainability and environmental impact of local public finance decisions.

%\textbf{(spiegare perche' si usano costi in conto capitale per i) e costi correnti per ii) e iii))}

The second set of data comes from the National Waste Register (\textit{Catasto Nazionale dei Rifiuti}), managed by the Italian Institute for Environmental Protection and Research (\textit{Istituto Superiore per la Protezione e la Ricerca Ambientale, or ISPRA}). Specifically, the data obtained from ISPRA include: (iv) the percentage of sorted waste; (v) (per capita) costs of treatment for sorted waste; (vi) (per capita) costs of transport and collection of unsorted waste; (vii) (per capita) costs of collection and transport of sorted waste; and (viii) (per capita) total costs of waste management. Table \ref{tab:variables} contains the list of the variables.

\begin{table}[htbp]
\centering
\adjustbox{max height=\dimexpr\textheight-4.5cm\relax, max width=\textwidth}{
\begin{tabular}{p{8cm} p{6cm} p{6cm}}
\toprule
\textbf{Variable Name} & \textbf{Description} & \textbf{Source} \\
\midrule

\textit{i) Capital Expend. for Sust. Dev. and Protection of the Territory and the Environment}  & Capital expenditures related to environmental protection, natural resource management, pollution control, waste and water infrastructure, and soil defense. Includes investments supporting the planning, coordination, and implementation of regional sustainable development and environmental policies (euros/inhabitant per year) & OpenBDAP\\\hline
\textit{ii) Current Expend. for the Integrated Water Service} & Cover the day-to-day costs of managing water supply, sewage, and wastewater treatment. These include maintenance, energy, personnel, and monitoring services. They ensure the regular and safe operation of water systems (euros/inhabitant per year) & Open Bilanci, OpenBDAP \\\hline
\textit{iii) Current Expend. on Green Urban Areas} & Including parks, public gardens, tree-lined streets, and related services. These expenditures cover routine maintenance, cleaning, irrigation, minor repairs, and personnel costs, supporting the preservation and accessibility of urban green spaces (euros/inhabitant per year) & Open Bilanci, OpenBDAP \\\hline
\textit{iv) Percentage of Sorted Waste} & Ratio between sorted waste collected an total municipal waste collected & National Waste Register (ISPRA) \\\hline
\textit{v) Costs of Treatment and Recovery for Sorted Waste} & Costs of treatment and recycling of sorted municipal waste (euros/inhabitant per year) & \\\hline
\textit{vi) Costs of Transport and Collection of Unsorted Waste} & Costs of collection and transport of unsorted municipal waste (euros/inhabitant per year)  & National Waste Register (ISPRA)\\\hline
\textit{vii) Costs of Transport and Collection of Sorted Waste} & Costs of collection and transport of sorted municipal waste (euros/inhabitant per year) & National Waste Register (ISPRA) \\\hline
\textit{viii) Total Costs of Waste Management} & Total costs of waste management and urban hygiene (euros/inhabitant per year) & National Waste Register (ISPRA) \\
\bottomrule
\end{tabular}}
\caption{Variables to measure the efficiency of environmental policies}
\label{tab:variables}
\end{table}

Municipalities are required to report the economic costs associated with integrated waste management services to ISPRA following Art.189 of the Legislative Decree No. 152/2006 (Environmental Code), via the annual submission of the Environmental Declaration Form  (\textit{Modello Unico di Dichiarazione ambientale or MUD}). These reported costs serve as the basis for setting local waste tariffs (TARI). Unlike budgeted municipal expenditures, the information on the costs collected by ISPRA is tied to the services financed by the TARI tax, thus directly paid by citizens, rather than funded through other sources (e.g., transfers from the State, regional authorities, the EU, or municipal budget surpluses).

We also collected data that will provide controls in the econometric analysis, namely: disposable income per capita at the municipal level (from the Italian Ministry of Economics and Finance), and municipal population (from ISPRA). In addition, we collected information on whether each municipality manages waste independently or is part of a consortium responsible for waste management services that will provide another control variable.

Data on the dissolution of infiltrated municipalities come from the Central Statistics Office of the Italian Ministry of Internal Affairs. Table \ref{tab:mun} shows the breakdown of municipalities dissolved per year from 2011 to 2022 in each Region.

\begin{table}[htp]
    \centering
   \adjustbox{max height=\dimexpr\textheight -4.5cm\relax,
           max width=\textwidth}{
\begin{tabular}{@{\extracolsep{5pt}} cccccccccccccccc} 
\\[-1.8ex]\hline 
\hline \\[-1.8ex] 
Region & 2010 & 2011 & 2012 & 2013 & 2014 & 2015 & 2016 & 2017 & 2018 & 2019 & 2020 & 2021 & 2022 & Total & \% of Total\\ 
\hline \\[-1.8ex] 
Basilicata &  &  &  &  &  &  &  &  &  & 1 &  &  &  & 1 & 1\% \\ 
Calabria & 3 & 4 & 9 & 8 & 5 & 3 & 3 & 12 & 11 & 6 & 4 & 4 & 3 & 75 & 44\% \\ 
Campania &  &  & 6 & 3 & 1 & 1 & 2 & 4 & 3 & 2 & 2 & 2 & 4 & 30 & 18\% \\ 
Emilia-Romagna &  &  &  &  &  &  & 1 &  &  &  &  &  &  & 1 & 1\% \\ 
Lazio &  &  &  &  &  & 1 &  &  &  &  &  &  & 2 & 3 & 2\% \\ 
Liguria &  &  &  &  &  &  &  & 1 &  &  &  &  &  & 1 & 1\% \\ 
Lombardia &  &  &  & 1 &  &  &  &  &  &  &  &  &  & 1 & 1\% \\ 
Piemonte &  &  & 2 &  &  &  &  &  &  &  &  &  &  & 2 & 1\% \\ 
Puglia &  &  &  &  & 1 & 1 &  & 2 & 4 & 3 & 1 & 4 & 2 & 18 & 11\% \\ 
Sicilia &  & 1 & 5 & 3 & 2 & 2 & 2 & 2 & 5 & 7 & 3 & 4 &  & 36 & 21\% \\ 
Valle d'Aosta &  &  &  &  &  &  &  &  &  &  & 1 &  &  & 1 & 1\% \\ 
\bottomrule\\[-1.8ex]
\end{tabular}}
    \caption{\emph{Number of Dissolved Municipalities from 2010 to 2022}}
    \label{tab:mun}
\end{table}

As shown in Figure \ref{fig:dissolved}, Table \ref{tab:mun}, highlight that the dissolution of infiltrated municipalities also affected, albeit with lower intensity, Italian regions located in the Center-North, such as Emilia-Romagna, Lazio, Liguria, Lombardia, Piemonte and Valle d'Aosta.

%\textbf{Va messo il dettaglio della fonte dei dati sugli scioglimenti, e tenuto pronto un breakdown degli scioglimenti in Campania, Calabria, Puglia, Sicilia vs il totale degli scioglimenti}

\subsection{Methodolgy\label{secMethod}}
We exploited the timing of the dismissal of municipalities for mafia infiltration from 2010 to 2022 to assess the effect of mafia infiltration on the environmental performance of these municipalities. Our conjecture is that mafia infiltration causally contributes to reduce the efficiency of environmental policies, so that the efficiency will improve after the dissolution of the city council. Our assumption is that an infiltrated city council is less efficient in implementing environmental policies for at least two reasons: on the one hand, a criminal organization can benefit from inefficient environmental policies as it can directly profit from them. For example, a criminal organization can provide illegal disposal of toxic waste as long as the service provided by local institutions is inefficient, or it can benefit from inefficient waste sorting, as long as it can benefit from the disposal in landfills under its control \citep[p. 186]{d2015waste}. In addition, because an infiltrated council is less efficient in general. Existing evidence suggests that, indeed, criminal organizations reduce the human capital level of elected politicians \citep{baraldi2022self}, which can reduce the efficiency of any council initiative.\footnote{Inefficiency of official institutions is one of the reasons that can increase the social consensus for organized crime \citep{lavezzi2014organised}. For this reason, organized crime has an incentive to reduce the level of institutional efficiency.}

The first step to identify the causal effect of the dissolution of mafia-connected municipalities on their environmental performances is to estimate a two-way fixed effect (TWFE) model. To identify the average treatment effect, first of all, we estimate the following specification:
\begin{equation}
Y_{i_t} = \beta_0+\beta_1 Law164_{i_t} +\gamma \mathbf{X}_{i_t} +\alpha_i+\delta_t+\epsilon_{i_t} 
\label{eqATE}
\end{equation}
where $Y_{i_t}$ is one of the variables i)-viii) for municipality $i$ at time $t$; $Law164_{i_t}$ is a dummy variable that takes the value 1 in the periods after municipality $t$ has been dissolved and 0 otherwise. To take into account the period after dissolution in which the municipality is administered by a commission, we will consider two alternative: one in which the ``treatment" starts the year following the dissolution, and one in which it starts after 3 years. Considering three years after dissolution is a way to fully consider the period available to the Commission to implement a new expenditure policy.\footnote{This choice in the same spirit of \citet[Table 3]{galletta2017law} who assesses the drop in the municipal public expenditure after dissolution in the year immediately following the intervention and in a three-year period after the intervention.}  $\mathbf{X}_{i,t}$ is a vector of controls and $\gamma$ is the vector of their coefficients; $\alpha_i$ and \(\delta_t\) represent, respectively, municipal and time fixed effects; $\epsilon_{i,t}$ is a white-noise disturbance.

The coefficient of interest is $\beta$ that represents the effect of \(Law164_{i_t}\) on the environmental performance of dissolved municipalities due to mafia infiltration. The model isolates the within-unit variation over time by including municipal (and provincial in some specifications) fixed effects $ (\alpha_i)$, helping to control for unobserved, time-invariant characteristics that could correlate with the treatment and the outcome. Time fixed effect $\delta_t$ address any systematic temporal influences that affect all municipalities similarly but are unrelated to the treatment itself, such as national economic cycles or external shocks.

The next step of the empirical strategy consists in an event-study analysis, to identify the effect of the dissolution in the periods after the implementation of the policy, by comparing it to the dynamics of the variable of interest before the event. To this purpose, we resort to the staggered difference-in-differences (DiD) approach. Staggered DiD extends the standard DiD method to settings where the treatment is implemented at different times across units rather than simultaneously. The staggered DiD model allows for a more nuanced analysis by incorporating variation in treatment timing, thus helping to capture heterogeneous treatment effects across different units and time periods \citep{CallawaySantAnna2021}.

More in detail, the specification we consider is:

\begin{equation}
Y_{i_t} = \sum_{n=-3}^{+7} \upsilon_i*D_t +\alpha_i+\delta_t+\gamma \mathbf{X}_{i_t}+\epsilon_{i_t}
\label{eqEvent}
\end{equation}
where \(D_t\) is the set of event-time dummies taking the value 1 after the dissolution of the municipality (one year or three years later). For the purposes of this article, we consider an interval between three years before and seven years after the dissolution. In particular, we consider a fairly high number of years after the dissolution as, on the one hand, the implementation of novel environmental policies might require some years for its setup, for example for new investments, and, on the other hand, some years might be necessary to see the outcomes of the new policies.

One of the key assumptions in staggered DiD is the parallel trends assumption for each cohort: in the absence of treatment, treated and untreated units within each cohort would follow similar trends over time. Furthermore, recent research has emphasized the importance of properly weighting each cohort to avoid biases that can arise from the different timing of treatment \citep{GoodmanBacon2021}.

In the event study estimation we follow \citet{borusyak2024revisiting}. This approach leads to an estimator that explicitly incorporates the researcher’s goal and assumptions about parallel trends, anticipation effects, and restrictions on treatment-effect heterogeneity. It is constructed by estimating a flexible high-dimensional regression that differs from conventional event study specifications and aggregating its coefficients appropriately. This estimator more generally ensures unbiasedness and yields attractive efficiency properties. The estimator is implemented through an ``imputation'' procedure.\footnote{This imputation follows the following steps \citep{borusyak2024revisiting}:
i) municipal and time fixed effects $\hat{\alpha_i}$ and $\hat{\beta_i}$ are fitted by regressions using untreated observations only; ii) these fixed effects are used to impute the untreated potential outcomes and therefore obtain an estimated treatment effect $\hat{\tau_{i_t}}$ = $\hat{Y_{i_t}}-\hat{\alpha_i}-\hat{\beta_t}$ for each treated observation; iii) a weighted sum of these treatment-effect estimates is taken, with weights corresponding to the estimation target, namely an average of some heterogeneous causal effect between the treatment and the outcome variable.}

Briefly, denoting $Y_{i_t}(0)$ the period-$t$ stochastic potential outcome of unit $i$ if it is never treated, causal effects on the treated observations $i_t \in \Omega_1$ are denoted $\tau_{i_t} = E [Y_{i_t} - Y_{i_t}(0)]$. Considering a statistic which sums or averages treatment effects $\tau = (\tau_{i_t})_{it\in\Omega_1}$ over
the set of treated observations with pre-specified non-stochastic weights $w_1 = (w_{i_t})_{it\in\Omega_1}$ that can depend on treatment assignment and timing, but not on realized outcomes, the estimation target can be defined as:

\begin{equation}
 \tau_w = \sum_{it\in\Omega_1} w_{i_t}\tau_{i_t} \equiv w'\tau
 \label{esTarg}
\end{equation}

The weights can be selected according to the research question. In the case of this analysis, we used this approach for the event study, setting the weights as the average effect $h$ periods since treatment for a given horizon $h \geq 0$. Thus, $w_{i_t} = 1 [K_{i_t} = h] / \big|\Omega_1,h|$ for $\Omega_{1,h} = {i_t: K_{i_t} = h}$ where $K_{i_t} = t - E_i$ is the number of periods since the event $E_i$ date, that we denote the ``horizon''. 

\section{Results\label{secResults}}
As mentioned, we estimate the Average Treatment Effect on the Treated (ATT) in two cases: i) assuming the treatment starts from the first year the municipality is dissolved; ii) assuming that the treatment starts three years after the dissolution, that is after the period the municipality has been administered by an external commission, so that the effects of its activity can be evaluated after the available time to work is elapsed. %\textbf{(I casi considerati da Galletta (vedi sua Tab. 3) sono: i) una dummy =1 l'anno del dissolvimento, 0 altrimenti; ii) una dummy = 1 l'anno del dissolvimento e i due anni successivi). Idem quando considera il caso degli spillovers tra comuni contigui. Controllare)}
%This allows us to compare the effects of the dissolution immediately after the dissolution itself and when a new municipal council has been reintroduced. In addition, an intervention in municipal environmental policy is likely to require some time to be set up, so it is worth considering the possibility that the "treatment" does not start at the onset of the commissioner's period as head of the administration, but rather some years afterwards. 

We run two different specifications: one with municipal and time fixed effects only, and the other including controls. The first set of controls for all outcome variables includes population and income per capita. In addition, for the variables related to waste management, we also use a dummy variable that indicates whether the municipality is included within a consortium or not.  All specifications include standard errors clustered at the municipal level. 

Tables \ref{tab:munexp} and \ref{tab:wman} present the results for, respectively, the measures indicated above as i)-iii) and iv)-viii).

%\textbf{(va spiegata bene la differenza tra le voci usate da Di Cataldo e Mastrorocco su ``waste'' e quello che facciamo qui, oltre a spiegare bene la differenza, se c'è, tra spese correnti e in conto capitale e tra queste e i costi del servizio. I costi per sviluppo sostenibile ecc, cioe' la voce i), come si collegano ad esempio ai costi per rifituti differenziati, voci v) e vii) e ai costi totali (voce viii)?)}

% \textwidth
\begin{table}[H]
\caption{Municipal Expenditures (variables i) - iii))}
\label{tab:munexp}
\centering
\adjustbox{max height=\dimexpr\textheight-10cm\relax, max width=\textwidth}{
\begingroup
%\renewcommand{\arraystretch}{1.2}
 % Prima sottotabella
        \begin{tabular}{lcccc}
        \toprule
        \multicolumn{5}{c}{\textit{i) Capital Expenditures on Sustainable Development and the Environment}}\\
        \midrule
        & \multicolumn{2}{c}{\textit{Treatment starts after 3 years}} & \multicolumn{2}{c}{\textit{Treatment starts after dissolution}} \\
        \midrule\midrule
        \emph{Variables} \\
        Law164          & 0.4003$^{***}$ & 0.3622$^{***}$ & 0.4848$^{***}$ & 0.4421$^{***}$ \\   
                        & (0.1111)       & (0.1079)       & (0.1323)       & (0.1313) \\   
        \midrule
        \emph{Fit statistics} \\
        Observations   & 48,969 & 48,158 & 48,969 & 48,158 \\  
        R$^2$          & 0.27934 & 0.29335 & 0.27944 & 0.29344 \\  
        Within R$^2$   & 0.00034 & 0.01754 & 0.00047 & 0.01766 \\  
        \midrule
         \multicolumn{5}{c}{\textit{ii) Current Expenditures on Integrated Water Service}}\\
        \midrule
        & \multicolumn{2}{c}{\textit{Treatment starts after 3 years}} & \multicolumn{2}{c}{\textit{Treatment starts after dissolution}} \\
\midrule\midrule
\emph{Variables} \\
        Law164      & 0.5388$^{***}$ & 0.5114$^{***}$ & 0.4162$^{***}$ & 0.3862$^{***}$ \\   
                      & (0.1263)       & (0.1264)       & (0.1131)       & (0.1133) \\   
        \midrule
        \emph{Fit statistics} \\
        Observations  & 79,674 & 78,801 & 79,674 & 78,801 \\  
        R$^2$         & 0.70025 & 0.69392 & 0.70018 & 0.69385 \\  
        Within R$^2$  & 0.00071 & 0.00233 & 0.00048 & 0.00210 \\
        \midrule
        \multicolumn{5}{c}{\textit{iii) Current Expenditures on Green Urban Areas}}\\
        \midrule
        & \multicolumn{2}{c}{\textit{Treatment starts after 3 years}} & \multicolumn{2}{c}{\textit{Treatment starts after dissolution}} \\
   \midrule\midrule
   \emph{Variables}\\
   Law164          & 0.1683                & 0.1225 & 0.1784                  & 0.1293\\   
                      & (0.1248)              & (0.1244) & (0.1146)              & (0.1146)\\   
   \midrule
   \emph{Fit statistics}\\
   Observations       & 79,675                & 78,802  & 79,675                & 78,802\\  
   R$^2$              & 0.59807               & 0.59777 & 0.59807               & 0.59777\\  
   Within R$^2$       & $8.49\times 10^{-5}$  & 0.00306 $6.69\times 10^{-5}$  & 0.00305\\  
   
        \midrule
        Municipality FE         & Yes & Yes & Yes & Yes \\  
        Year FE          & Yes & Yes & Yes & Yes \\ 
        Controls & No & Yes & No & Yes\\
       \bottomrule
        \multicolumn{4}{l}{\emph{Clustered (Municipality) standard-errors in parentheses}}\\
   \multicolumn{4}{l}{\emph{Signif. Codes: ***: 0.01, **: 0.05, *: 0.1}}\\
        \end{tabular}\endgroup
        }
\end{table}

\begin{table}[H]
\vspace{-3cm}
\caption{Waste Management Costs (Variables iv) - viii))}
\label{tab:wman}
\centering
\adjustbox{max height=\dimexpr\textheight-10cm\relax, max width=\textwidth}{
\begingroup
%\renewcommand{\arraystretch}{1.2}
 % Prima sottotabella
        \begin{tabular}{lcccc}
        \toprule
        \multicolumn{5}{c}{\textit{iv) Percentage of Sorted Waste}}\\
        \midrule
        & \multicolumn{2}{c}{\textit{Treatment starts after 3 years}} & \multicolumn{2}{c}{\textit{Treatment starts after dissolution}}\\
   \midrule \midrule
   \emph{Variables}\\
   Law164          & 0.0715$^{***}$ & 0.0661$^{***} $& 0.0453$^{***}$ & 0.0384$^{**}$\\   
                      & (0.0166)       & (0.0166) & (0.0149)       & (0.0150)\\   
   \midrule

        \emph{Fit statistics}\\
   Observations       & 80,057         & 78,826  & 80,057         & 78,826\\  
   R$^2$              & 0.79743        & 0.79782 & 0.79729        & 0.79769\\  
   Within R$^2$       & 0.00128        & 0.00500 & 0.00059        & 0.00432\\  
   \midrule
         \multicolumn{5}{c}{\textit{v) Costs of Treatment of Sorted Waste}}\\
        \midrule
        & \multicolumn{2}{c}{\textit{Treatment starts after 3 years}} & \multicolumn{2}{c}{\textit{Treatment starts after dissolution}}\\
        \midrule\midrule
        \emph{Variables}\\
   Law164          & 0.3602$^{***}$ & 0.3451$^{***}$ & 0.2828$^{**}$ & 0.2693$^{**}$\\   
                      & (0.1146)       & (0.1145) & (0.1102)      & (0.1100)\\   
   \midrule
\emph{Fit statistics}\\
   Observations       & 45,898         & 44,838 & 45,898        & 44,838\\  
   R$^2$              & 0.58306        & 0.58196 & 0.58295      & 0.58186\\  
   Within R$^2$       & 0.00076        & 0.00402 & 0.00050      & 0.00378\\  
   \midrule
        \multicolumn{5}{c}{\textit{vi) Costs of Collection and Transport of Unsorted Waste}}\\
        \midrule
        & \multicolumn{2}{c}{\textit{Treatment starts after 3 years}} & \multicolumn{2}{c}{\textit{Treatment starts after dissolution}}\\
   \midrule\midrule
   \emph{Variables}\\
   Law164          & -0.2266$^{***}$ & -0.2271$^{***}$ & -0.2074$^{**}$ & -0.2095$^{**}$\\   
                      & (0.0790)        & (0.0790)  & (0.0924)      & (0.0924)\\   
   \midrule
   \emph{Fit statistics}\\
   Observations       & 54,158          & 52,985  & 54,158         & 52,985\\  
   R$^2$              & 0.66662         & 0.66684 & 0.66659        & 0.66681 \\  
   Within R$^2$       & 0.00043         & 0.00128 & 0.00032        & 0.00118\\  
\midrule
\multicolumn{5}{c}{\textit{vii) Costs of Collection and Transport of Sorted Waste}}\\
\midrule
& \multicolumn{2}{c}{\textit{Treatment starts after 3 years}} & \multicolumn{2}{c}{\textit{Treatment starts after dissolution}}\\
   \midrule\midrule
   \emph{Variables}\\
   Law164          & 0.2866$^{***}$ & 0.2934$^{***}$ & 0.2217$^{**}$ & 0.2934$^{***}$\\   
                      & (0.1064)       & (0.1075) & (0.1039)      & (0.1075)\\   
   \midrule
   \emph{Fit statistics}\\
   Observations       & 53,786         & 52,613 & 53,786        & 52,613\\  
   R$^2$              & 0.60257        & 0.59721 & 0.60250       & 0.59721\\  
   Within R$^2$       & 0.00052        & 0.00352 & 0.00034       & 0.00352\\  
   \midrule
   \multicolumn{5}{c}{\textit{viii) Total Costs of Waste Management}}\\
   \midrule
  & \multicolumn{2}{c}{\textit{Treatment starts after 3 years}} & \multicolumn{2}{c}{\textit{Treatment starts after dissolution}}\\
   \midrule\midrule
   \emph{Variables}\\
   Law164          & 0.0571$^{**}$ & 0.0614$^{**}$ & 0.0344           & 0.0369\\   
                      & (0.0270)      & (0.0272) & (0.0255)              & (0.0256)\\   
  \midrule
   \emph{Fit statistics}\\
   Observations       & 55,744                & 54,559 & 55,744        & 54,559\\  
   R$^2$              & 0.85057               & 0.84736 & 0.85059       & 0.84739\\  
   Within R$^2$       & $8.67\times 10^{-5}$  & 0.00356 & 0.00022       & 0.00371\\  
   \midrule
         Municipality FE         & Yes & Yes & Yes & Yes \\  
        Year FE          & Yes & Yes & Yes & Yes \\ 
        Controls & No & Yes & No & Yes\\
        \bottomrule
        \multicolumn{4}{l}{\emph{Clustered (Municipality) standard-errors in parentheses}}\\
   \multicolumn{4}{l}{\emph{Signif. Codes: ***: 0.01, **: 0.05, *: 0.1}}\\
        \end{tabular}\endgroup
        }
\end{table}

The results in Table \ref{tab:munexp} for the measures i)-iii), that is for indicators of environmental expenditures, show that, with the exception of the expenditure for urban areas (measure iii)), there is a significant and positive effect coming from the dissolution of the infiltrated council. The impact is detected assuming that the treatment starts the year of the dissolution as well as in the case the treatment is assumed to start after three years.

The results for measures iv)-viii) (Table \ref{tab:wman} ) show that: the percentage of sorted waste significantly increases; the cost for the treatment and for the collection and transport of sorted waste increases, while the costs related to unsorted waste decrease. Overall, the total cost of waste management slightly increases after three years from the dissolution, as a result of the two opposing tendencies identified for sorted and unsorted waste.

Together, these results suggest that after dissolution the environmental policies improve as: i) there is higher expenditure in environmental policies, in particular in capital expenditure for sustainable development and the environment, and in the integrated water service; ii) the percentage of sorted waste increases (a result in line with \citealp{baraldi2024fighting}); iii) the costs of sorted waste management increase (in both components: treatment,  collection and transport) and the costs for unsorted waste management decreases. This represents a clear shift in the composition of the public expenditure that gives higher priority to sorted waste than to unsorted waste, which can be considered as an allocation of public expenditure less favorable to the interests of organized crime. iv) the total cost of waste management increases, in particular when we assume that the treatment acts after three years of the dissolution, probably as the effect of the increase the costs for sorted waste management prevails over the decrease in the costs for unsorted waste management when we consider the whole period in which the commission had the possibility to work.

These results provide a different perspective than that of \citet{di2022organized} who find that the share of public expenditure on Construction and Waste Management, sectors assumed to be more favorable for infiltration by criminal organizations, decreases after dissolution. \citet{di2022organized}, however, consider the expenditure for waste management together with the expenditure for Construction (roads, infrastructure, etc) and do not distinguish between expenditures for the management of sorted and unsorted waste as we do here. In addition, as noted, we find that the total costs for waste management increase after the dissolution, due to the increase in the costs of sorted waste management. The expenditure data used by \citet{di2022organized} are derived from budgeted capital expenditures, which refer to long-term investments financed through various sources of municipal revenue. In the case of waste management, these budgeted expenditures typically include costs related to waste management, as well as street cleaning and environmental awareness campaigns. In contrast, the cost data used in this study, collected from ISPRA, are directly related to the waste management service and are exclusively funded through the waste tariff (TARI). Unlike budgeted expenditures, the ISPRA data used in this analysis allow for a more detailed assessment of the specific costs associated with different waste treatment options (e.g., sorted vs. unsorted waste) and the distinct phases of the waste management process (e.g., transport, collection, and treatment). These results complement those of  \citet{baraldi2024fighting}, providing a more detailed analysis of the change in waste management policy. In this perspective, the increase in the total waste management costs that we detect when considering three years after the dissolution, should not be considered as wasteful expenditures, but reflect a shift towards a more environmental-conscious policy, which becomes evident thanks to the breakdown of the costs in the components for sorted and unsorted waste.

\subsection{Event Study}
To examine the behavior of the variables accounting for environmental policy before and after the council dissolution, that we considered as a staggered treatment, we applied the methodology of \citet{borusyak2024revisiting} to our data on a time span going from 3 years before the dissolution to 7 years after. This choice allows us to consider a fairly large period after dissolution, which might be necessary for the environmental policy changes to display their effects. In this event study we keep municipal and year fixed effects and consider the variables having significant and robust results from the estimation of Eq. (\ref{eqATE}), leaving out the variable ``Current Expenditures on Green Urban Areas" (variable iii)). 

\begin{figure}[H]
    \centering
\includegraphics[width=0.9\linewidth]{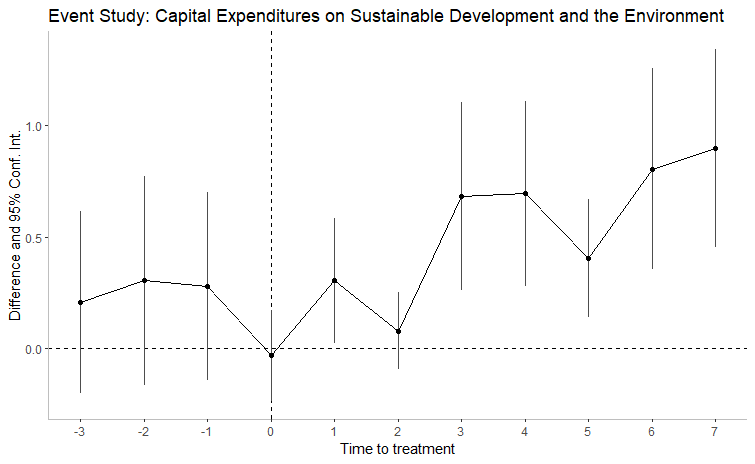}
    \caption{Event study: i) Capital Expenditures Sustainable on Development and Protection of Territory and the Environment}
    \label{fig:f9}
\end{figure}

\begin{figure}[H]
    \centering
\includegraphics[width=0.9\linewidth]{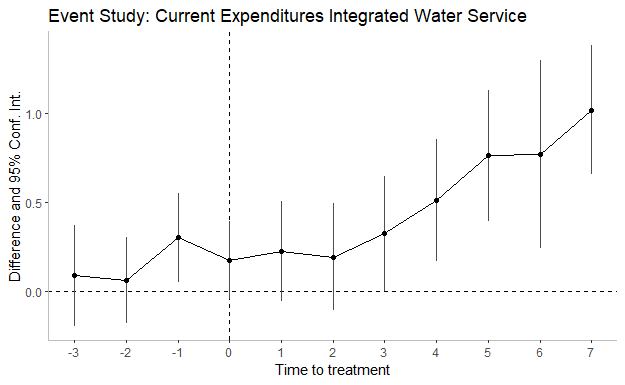}
    \caption{Event study: ii) Current Expenditures Integrated Water Service}
    \label{fig:cod4}
\end{figure}

\begin{figure}[H]
    \centering
\includegraphics[width=0.9\linewidth]{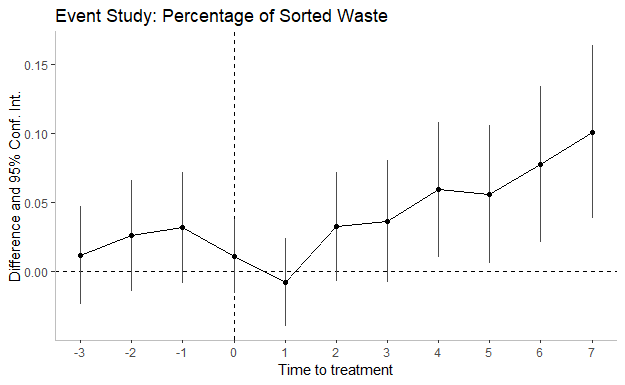}
    \caption{Event study: iv) Percentage of sorted waste}
    \label{fig:rd}
\end{figure}

\begin{figure}[H]
    \centering
\includegraphics[width=0.9\linewidth]{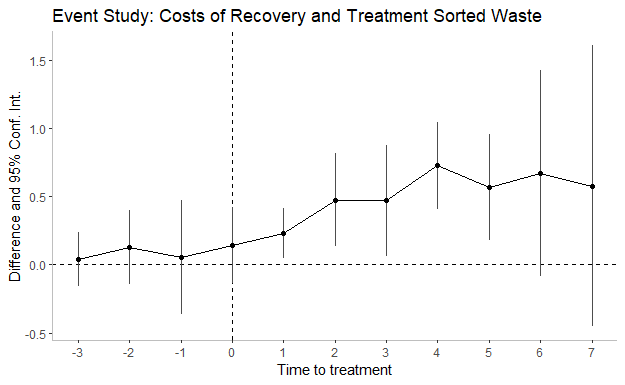}
    \caption{Event study: v) Cost of recovery and treatment of sorted waste}
    \label{fig:costtreatment}
\end{figure}

\begin{figure}[H]
    \centering
\includegraphics[width=0.9\linewidth]{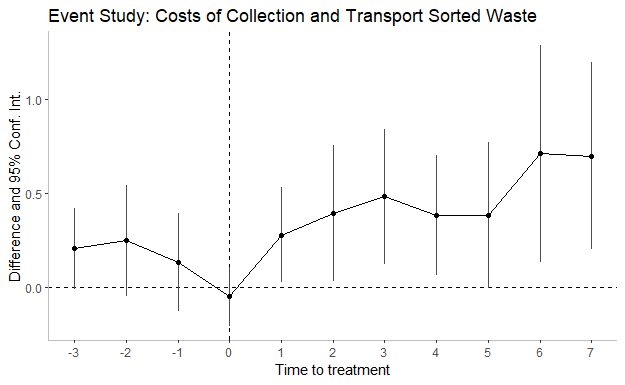}
    \caption{Event study: vi) Costs of collection and transport of sorted waste}
    \label{fig:sortcosts}
\end{figure}

\begin{figure}[H]
    \centering
\includegraphics[width=0.9\linewidth]{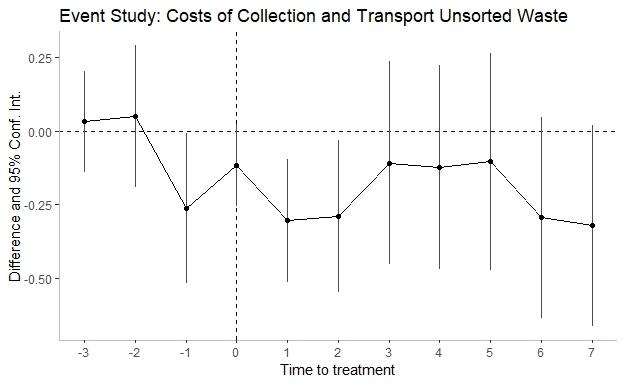}
    \caption{Event study: vii) Costs of collection and transport of unsorted waste}
    \label{fig:unsortcosts}
\end{figure}

\begin{figure}[H]
    \centering
\includegraphics[width=0.9\linewidth]{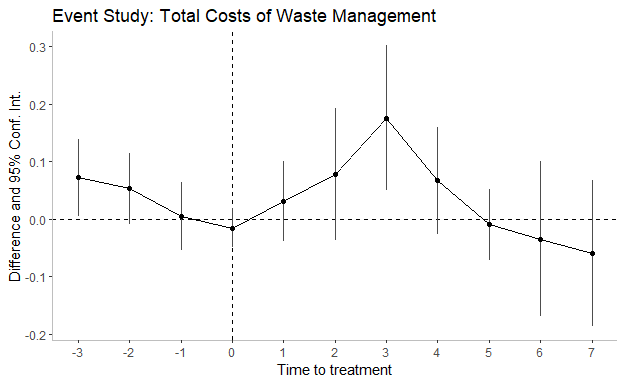}
    \caption{Event study: viii) Total Costs of Waste Management}
    \label{fig:totcost}
\end{figure}

Figures \ref{fig:f9}-\ref{fig:totcost} show the results. From a visual analysis of the different graphs, all of the variables broadly respect the parallel trend hypothesis, that is, with few exceptions the estimated coefficient for the time dummy in Eq. (\ref{eqEvent}) for the periods before the dissolution is not statistically different from zero. 

Figure \ref{fig:f9} shows the event study of Capital Expenditures, right immediately after dissolution the coefficient starts increasing, keeping this trend up to the end of the period considered. The same behavior is shown for Current Expenditures of the Integrated Water Service (Figure \ref{fig:cod4}) with the coefficient starting to increase after two years from the dissolution.  

%\subsubsection{Waste Management}

Figure \ref{fig:rd} shows the behavior before and after the dissolution for the percentage of collected sorted waste. Some years after the dissolution a significant positive trend clearly appears. When observing the dynamics of the costs directly related to sorted waste, recovery and treatment (Fig. \ref{fig:costtreatment}) and collection and transport (Fig. \ref{fig:sortcosts}), we notice that the positive trend appears right after the dissolution of the city council (and appears to fade for the costs of Recovery and Treatment after six years from the dissolution). These results, taken together, suggest that after the dissolution more resources are devoted to the management of sorted waste (visible from the increase in the sorted waste management costs), which after some years causes a significant increase in the percentage of sorted waste. These results are in line with those of \cite{baraldi2024fighting} with respect to the dynamics of the percentage of sorted waste. \cite{baraldi2024fighting}, however, do not consider the dynamics of the related management costs as we do in this article.
 
%Findings based on a staggered differences-in-differences estimations indicate that expenditures in integrated water service increase right after the city council dissolution (Figure \ref{fig:Cod4}). With respect to the share sorted waste, Mafia infiltration has a negative effect that ceases after the dissolution, and tends to increase in the following years (Figure \ref{fig:rd}). Finally, unsorted waste management costs tend to decrease with the dissolution of the city council (Figure  \ref{fig:Unscost}).

As regards the costs of collection and transport of unsorted waste (Figure \ref{fig:unsortcosts}), the trend appears negative right after the dissolution and for most of the subsequent periods, confirming that there is a clear change of policy after the dissolution with more resources devoted to sorted waste.

Finally, the total costs of waste management display an increasing trend after the dissolution, which disappears around the fourth year (Figure \ref{fig:totcost}). As noted above, this likely reflects the joint effect of the increase in the costs of sorted waste and the decrease in costs of unsorted waste, signaling the change in priorities of the new administration toward more sustainable waste management processes. Eventually, costs for the municipality start to reduce several years after the dissolution.  

%Results thus contribute to the existing literature \citep{d2015waste,baraldi2022self}, providing insights into the efficiency patterns of the waste management system when the link between organized crime and local administration has been severed. Our findings show that the overall costs of waste management increase as a results of an increase in the costs of treatment and collection of sorted waste. This is the product of 

%\subsection{Robustness Check: Councils Dismissal for other causes}
%Other than mafia infiltration, municipalities can be dissolved for other causes, such as the impossibility of approving the annual budget, the resignation of the mayor or the majority of the municipal council, and the death of the mayor (Art 141 legislative decree 267/2000). We thus analyzed the ATT of those causes of dissolution on our outcome variable as a robustness check of our analysis. Tables ..., shows the results of the ATT for all the outcome variables considered. Most of them show an opposite trend with respect to the results of council dissolution due to mafia infiltration. Especially the ones on Sorted Waste  (Table ...) and the related costs of treatment and recovery (Table ...)

\subsection{Robustness check}
In this section, we present the results of some robustness checks. First of all, for this kind of analysis, the choice of the control group is crucial. For example, a municipality can be dissolved for other reasons than infiltration by organized crime. According to Art. 141 of the Legislative Decree 267/2000,  municipalities can be dissolved for the impossibility of approving the annual budget, for the resignation of the mayor or of the majority of the municipal council, and for the death of the mayor. To take this into account, we run the ATT analysis removing from the sample those municipalities that experienced dissolution for other causes in the period of observation (see also \citealp[p. 94]{galletta2017law}).

Table \ref{munexp:nomafia} and Table \ref{wman:nomafia} show the results for this specification. Our main results do not seem to be affected by the inclusion of municipalities whose city councils were dissolved for other reasons.

\begin{table}[H]
\caption{Municipal Expenditures (variables i)- ii). Municipalities dissolved for other causes are excluded}
\label{munexp:nomafia}
\centering
\adjustbox{max height=\dimexpr\textheight-10cm\relax, max width=14cm}{
\begingroup
%\renewcommand{\arraystretch}{1.2}
 % Prima sottotabella
        \begin{tabular}{lcccc}
        \toprule
        \multicolumn{5}{c}{\textit{i) Capital Expenditures Sustainable Development of the Territory and the Environment}}\\
        \midrule
        & \multicolumn{2}{c}{\textit{Treatment starts after 3 years}} & \multicolumn{2}{c}{\textit{Treatment starts after dissolution}}\\
        \midrule\midrule
        \emph{Variables} \\
        Law164             & 0.5425$^{***}$ & 0.5055$^{***}$ & 0.4054$^{***}$ & 0.3766$^{***}$\\   
                      & (0.1915)       & (0.1916) & (0.1479)             & (0.1452)\\      
        \midrule
        \emph{Fit statistics}\\
   Observations       & 40,516         & 39,719  & 40,516         & 39,719\\  
   R$^2$              & 0.28329        & 0.29758 & 0.28318        & 0.29747\\  
   Within R$^2$       & 0.00040        & 0.01761 & 0.00023        & 0.01746\\  
 
        \midrule
         \multicolumn{5}{c}{\textit{ii) Current Expenditures Integrated Water Service}}\\
        \midrule
        & \multicolumn{2}{c}{\textit{Treatment starts after 3 years}} & \multicolumn{2}{c}{\textit{Treatment starts after dissolution}}\\
\midrule\midrule
\emph{Variables} \\
        Law164          & 0.6573$^{***}$ & 0.6302$^{***}$ & 0.4291$^{***}$ & 0.3983$^{***}$\\   
                      & (0.1426)       & (0.1426)       & (0.1306)       & (0.1308)\\   
      
        \midrule
        \emph{Fit statistics}\\
   Observations       & 66,239         & 65,385  & 66,239         & 65,385\\  
   R$^2$              & 0.70158        & 0.69357 & 0.70148        & 0.69347\\  
   Within R$^2$       & 0.00069        & 0.00230 & 0.00034        & 0.00196\\  
        \midrule
        \multicolumn{5}{c}{\textit{iii) Current Expenditures Green Urban Areas}}\\
        \midrule
        & \multicolumn{2}{c}{\textit{Treatment starts after 3 years}} & \multicolumn{2}{c}{\textit{Treatment starts after dissolution}}\\
   \midrule\midrule
   \emph{Variables}\\
   Law164             & 0.2394                & 0.1993         & 0.2237                & 0.1800\\   
                      & (0.1606)              & (0.1602)       & (0.1492)              & (0.1495)\\      
   \midrule
   \emph{Fit statistics}\\
   Observations       & 66,240                & 65,386 & 66,240                & 65,386\\  
   R$^2$              & 0.59853               & 0.59775 & 0.59853              & 0.59775\\  
   Within R$^2$       & $8.82\times 10^{-5}$  & 0.00253 & $8.78\times 10^{-5}$  & 0.00252\\  
   \midrule
        \midrule
        Municipality FE         & Yes & Yes & Yes & Yes \\  
        Year FE          & Yes & Yes & Yes & Yes \\ 
        Controls & No & Yes & No & Yes\\
       \bottomrule
        \multicolumn{4}{l}{\emph{Clustered (Municipality) standard-errors in parentheses}}\\
   \multicolumn{4}{l}{\emph{Signif. Codes: ***: 0.01, **: 0.05, *: 0.1}}\\
        \end{tabular}\endgroup
        }
\end{table}

\begin{table}[H]
\vspace{-3cm}
\caption{Waste Management Costs (Variables iv) - viii)). Municipalities dissolved for other causes are excluded}
\label{wman:nomafia}
\centering
\adjustbox{max height=\dimexpr\textheight-10cm\relax, max width=\textwidth}{
\begingroup
%\renewcommand{\arraystretch}{1.2}
 % Prima sottotabella
        \begin{tabular}{lcccc}
        \toprule
        \multicolumn{4}{c}{\textit{iv) Percentage of Sorted Waste}}\\
        \midrule
        & \multicolumn{2}{c}{\textit{Treatment starts after 3 years}} & \multicolumn{2}{c}{\textit{Treatment starts after dissolution}}\\
   \midrule \midrule
   \emph{Variables}\\
   Law164             & 0.1086$^{***}$ & 0.1036$^{***}$ & 0.0684$^{***}$ & 0.0615$^{***}$\\   
                      & (0.0238)       & (0.0238) & (0.0207)       & (0.0207)\\  \midrule

        \emph{Fit statistics}\\
   Observations       & 66,578         & 65,393 & 66,578         & 65,393\\  
   R$^2$              & 0.80070        & 0.80113 & 0.80049        & 0.80092\\  
   Within R$^2$       & 0.00197        & 0.00579 & 0.00090        & 0.00472\\  
   \midrule
  \multicolumn{4}{r}{\textit{v) Costs of Recovery and Treatment Sorted Waste}}\\
        \midrule
        & \multicolumn{2}{c}{\textit{Treatment starts after 3 years}} & \multicolumn{2}{c}{\textit{Treatment starts after dissolution}}\\
        \midrule\midrule
        \emph{Variables}\\
   Law164          & 0.4975$^{***}$ & 0.4820$^{***}$ & 0.3930$^{***}$ & 0.3812$^{**}$\\   
                      & (0.1452)       & (0.1437)       & (0.1521)       & (0.1510)\\      
   \midrule
\emph{Fit statistics}\\
   Observations       & 39,045         & 38,026  & 39,045         & 38,026\\  
   R$^2$              & 0.57734        & 0.57563 & 0.57724        & 0.57555\\  
   Within R$^2$       & 0.00089        & 0.00446 & 0.00067        & 0.00425\\  
   \midrule
        \multicolumn{4}{c}{\textit{vi) Costs of Collection and Transport Unsorted Waste}}\\
        \midrule
        & \multicolumn{2}{c}{\textit{Treatment starts after 3 years}} & \multicolumn{2}{c}{\textit{Treatment starts after dissolution}}\\
   \midrule\midrule
   \emph{Variables}\\
   Law164             & -0.2903$^{***}$ & -0.2907$^{***}$ & -0.2800$^{***}$ & -0.2772$^{***}$\\   
                      & (0.1121)        & (0.1122)        & (0.0951)        & (0.0954)\\    
   \midrule
   \emph{Fit statistics}\\
   Observations       & 45,643          & 44,511  & 45,643          & 44,511\\  
   R$^2$              & 0.66355         & 0.66373 & 0.66356         & 0.66374\\  
   Within R$^2$       & 0.00040         & 0.00128 & 0.00045         & 0.00132\\  
   \midrule
\multicolumn{4}{c}{vii) Costs of Collection and Transport Sorted Waste}\\
\midrule
& \multicolumn{2}{c}{\textit{Treatment starts after 3 years}} & \multicolumn{2}{c}{\textit{Treatment starts after dissolution}}\\
   \midrule\midrule
   \emph{Variables}\\
   Law164          & 0.4738$^{***}$ & 0.4848$^{***}$ & 0.3855$^{***}$ & 0.3932$^{***}$\\   
                      & (0.1421)       & (0.1438)       & (0.1195)       & (0.1199)\\    
   \midrule
   \emph{Fit statistics}\\
   Observations       & 45,308         & 44,173  & 45,308         & 44,173\\  
   R$^2$              & 0.60497        & 0.59852 & 0.60491        & 0.59846\\  
   Within R$^2$       & 0.00088        & 0.00422 & 0.00073        & 0.00405\\  
   \midrule
   \multicolumn{4}{c}{viii) Total Costs of Waste Management}\\
   \midrule
   & \multicolumn{2}{c}{\textit{Treatment starts after 3 years}} & \multicolumn{2}{c}{\textit{Treatment starts after dissolution}}\\
   \midrule\midrule
   \emph{Variables}\\
   Law164          & 0.0998$^{**}$ & 0.1032$^{**}$ & 0.0528   & 0.0534\\   
                      & (0.0413)      & (0.0418)      & (0.0380) & (0.0383)\\  
  \midrule
   \emph{Fit statistics}\\
   Observations       & 46,889        & 45,745  & 46,889   & 45,745\\  
   R$^2$              & 0.84858       & 0.84450 & 0.84854  & 0.84445\\  
   Within R$^2$       & 0.00041       & 0.00339 & 0.00014  & 0.00309\\  
    \midrule
         Municipality FE         & Yes & Yes & Yes & Yes \\  
        Year FE          & Yes & Yes & Yes & Yes \\ 
        Controls & No & Yes & No & Yes\\
        \bottomrule
        \multicolumn{4}{l}{\emph{Clustered (Municipality) standard-errors in parentheses}}\\
   \multicolumn{4}{l}{\emph{Signif. Codes: ***: 0.01, **: 0.05, *: 0.1}}\\
        \end{tabular}\endgroup
        }
\end{table}

Furthermore, the sample contains municipalities whose city council was dissolved more than once, suggesting that infiltration might have not been fully removed after the first dissolution (see \citealp[p. 10]{baraldi2024fighting}). For this reason, we re-estimate Eq. (\ref{eqATE}) excluding these municipalities. Looking at Table \ref{munexp:no2} and Table \ref{wman:no2} show that our results are robust to the exclusion of the municipalities with city councils dissolved multiple times in our period of observation.

\begin{table}[H]
\caption{Municipal Expenditures (variables i)- ii). Municipalities dissolved multiple times are excluded}
\label{munexp:no2}
\centering
\adjustbox{max height=\dimexpr\textheight-10cm\relax, max width=14cm}{
\begingroup
%\renewcommand{\arraystretch}{1.2}
 % Prima sottotabella
        \begin{tabular}{lcccc}
        \toprule
        \multicolumn{5}{c}{\textit{i) Capital Expenditures Sustainable Development of the Territory and the Environment}}\\
        \midrule
        & \multicolumn{2}{c}{\textit{Treatment starts after 3 years}} & \multicolumn{2}{c}{\textit{Treatment starts after dissolution}}\\
        \midrule\midrule
        \emph{Variables} \\
        Law164        & 0.4105$^{***}$ & 0.3713$^{***}$  & 0.3628$^{***}$ & 0.3277$^{***}$\\   
                      & (0.1417)       & (0.1411) 	 & (0.1155)       & (0.1122)\\      
        \midrule
        \emph{Fit statistics}\\
   Observations       & 48,885         & 48,074  & 48,885         & 48,074\\  
   R$^2$              & 0.27971        & 0.29367 & 0.27968        & 0.29364\\  
   Within R$^2$       & 0.00030        & 0.01744 & 0.00026        & 0.01740\\  
 
        \midrule
         \multicolumn{5}{r}{\textit{ii) Current Expenditures Integrated Water Service}}\\
        \midrule
        & \multicolumn{2}{c}{\textit{Treatment starts after 3 years}} & \multicolumn{2}{c}{\textit{Treatment starts after dissolution}}\\
\midrule\midrule
\emph{Variables} \\
        Law164        & 0.4992$^{***}$ & 0.4732$^{***}$ & 0.3665$^{***}$ & 0.3379$^{***}$\\   
                      & (0.1146)       & (0.1148)       & (0.1086)       & (0.1087)\\   
      
        \midrule
        \emph{Fit statistics}\\
   Observations       & 79,577         & 78,709   & 79,577        & 78,709\\  
   R$^2$              & 0.70024        & 0.69388 & 0.70018        & 0.69382\\  
   Within R$^2$       & 0.00056        & 0.00216 & 0.00035        & 0.00196\\  
        \midrule
        \multicolumn{5}{c}{\textit{iii) Current Expenditures Green Urban Areas}}\\
        \midrule
        & \multicolumn{2}{c}{\textit{Treatment starts after 3 years}} & \multicolumn{2}{c}{\textit{Treatment starts after dissolution}}\\
   \midrule\midrule
   \emph{Variables}\\
   Law164             & 0.1666                & 0.1222         & 0.1873                & 0.1396\\   
                      & (0.1306)              & (0.1300)       & (0.1173)              & (0.1171)\\      
   \midrule
   \emph{Fit statistics}\\
   Observations       & 79,578                & 78,710  & 79,578                & 78,710\\  
   R$^2$              & 0.59816               & 0.59788 & 0.59818               & 0.59789\\  
   Within R$^2$       & $5.98\times 10^{-5}$  & 0.00305 & $8.87\times 10^{-5}$  & 0.00307\\  
   \midrule
        \midrule
        Municipality FE         & Yes & Yes & Yes & Yes \\  
        Year FE          & Yes & Yes & Yes & Yes \\ 
        Controls & No & Yes & No & Yes\\
       \bottomrule
        \multicolumn{4}{l}{\emph{Clustered (Municipality) standard-errors in parentheses}}\\
   \multicolumn{4}{l}{\emph{Signif. Codes: ***: 0.01, **: 0.05, *: 0.1}}\\
        \end{tabular}\endgroup
        }
\end{table}

\begin{table}[H]
\vspace{-3cm}
\caption{Waste Management Costs (Variables iv) - viii)). Municipalities dissolved multiple times are excluded}
\label{wman:no2}
\centering
\adjustbox{max height=\dimexpr\textheight-10cm\relax, max width=\textwidth}{
\begingroup
%\renewcommand{\arraystretch}{1.2}
 % Prima sottotabella
        \begin{tabular}{lcccc}
        \toprule
        \multicolumn{4}{c}{\textit{iv) Percentage of Sorted Waste}}\\
        \midrule
        & \multicolumn{2}{c}{\textit{Treatment starts after 3 years}} & \multicolumn{2}{c}{\textit{Treatment starts after dissolution}}\\
   \midrule \midrule
   \emph{Variables}\\
   Law164             & 0.0817$^{***}$ & 0.0767$^{***}$ & 0.0528$^{***}$ & 0.0463$^{***}$\\   
                      & (0.0170)       & (0.0170) 	& (0.0151)       & (0.0151)\\  
\midrule

   \emph{Fit statistics}\\
   Observations       & 79,967         & 78,741  & 79,967         & 78,741\\  
   R$^2$              & 0.79725        & 0.79772 & 0.79709        & 0.79756\\  
   Within R$^2$       & 0.00154        & 0.00534 & 0.00076        & 0.00455\\  
   \midrule
  \multicolumn{4}{r}{\textit{v) Costs of Recovery and Treatment Sorted Waste}}\\
        \midrule
        & \multicolumn{2}{c}{\textit{Treatment starts after 3 years}} & \multicolumn{2}{c}{\textit{Treatment starts after dissolution}}\\
        \midrule\midrule
        \emph{Variables}\\
   Law164          & 0.3756$^{***}$ & 0.3602$^{***}$ & 0.2828$^{**}$ & 0.2692$^{**}$\\   
                   & (0.1164)       & (0.1162)       & (0.1102)      & (0.1100)\\      
   \midrule
\emph{Fit statistics}\\
   Observations       & 45,884         & 44,824  & 45,884         & 44,824\\  
   R$^2$              & 0.58307        & 0.58198 & 0.58294        & 0.58185\\  
   Within R$^2$       & 0.00080        & 0.00407 & 0.00050        & 0.00378\\  
   \midrule
        \multicolumn{4}{c}{\textit{vi)Costs of Collection and Transport Unsorted Waste}}\\
        \midrule
        & \multicolumn{2}{c}{\textit{Treatment starts after 3 years}} & \multicolumn{2}{c}{\textit{Treatment starts after dissolution}}\\
   \midrule\midrule
   \emph{Variables}\\
   Law164             & -0.2111$^{**}$ & -0.2138$^{**}$  & -0.2266$^{***}$ & -0.2272$^{***}$\\   
                      & (0.0955)       & (0.0955)        & (0.0790)        & (0.0790)\\    
   \midrule
   \emph{Fit statistics}\\
   Observations       & 54,136          & 52,963  & 54,136          & 52,963\\  
   R$^2$              & 0.66666         & 0.66688 & 0.66670         & 0.66692\\  
   Within R$^2$       & 0.00032         & 0.00118 & 0.00043         & 0.00128\\  
   \midrule
\multicolumn{4}{c}{vii) Costs of Collection and Transport Sorted Waste}\\
\midrule
& \multicolumn{2}{c}{\textit{Treatment starts after 3 years}} & \multicolumn{2}{c}{\textit{Treatment starts after dissolution}}\\
   \midrule\midrule
   \emph{Variables}\\
   Law164             & 0.3030$^{***}$ & 0.3113$^{***}$ & 0.2217$^{**}$ & 0.2278$^{**}$\\   
                      & (0.1098)       & (0.1110)       & (0.1039)      & (0.1044)\\    
   \midrule
   \emph{Fit statistics}\\
   Observations       & 53,763         & 52,590  & 53,763         & 52,590\\  
   R$^2$              & 0.60254        & 0.59719 & 0.60245        & 0.59709\\  
   Within R$^2$       & 0.00088        & 0.00422 & 0.00034        & 0.00334\\  
   \midrule
   \multicolumn{4}{c}{viii) Total Costs of Waste Management}\\
   \midrule
  & \multicolumn{2}{c}{\textit{Treatment starts after 3 years}} & \multicolumn{2}{c}{\textit{Treatment starts after dissolution}}\\
   \midrule\midrule
   \emph{Variables}\\
   Law164             & 0.0595$^{**}$ & 0.0641$^{**}$ & 0.0344   & 0.0369\\   
                      & (0.0278)      & (0.0280)      & (0.0255) & (0.0256)\\  
  \midrule
   \emph{Fit statistics}\\
   Observations       & 55,720        & 54,535  & 55,720   & 54,535\\  
   R$^2$              & 0.85059       & 0.84739 & 0.85057  & 0.84737\\  
   Within R$^2$       & 0.00023       & 0.00374 & $8.67\times 10^{-5}$  & 0.00357\\  
    \midrule
         Municipality FE         & Yes & Yes & Yes & Yes \\  
        Year FE          & Yes & Yes & Yes & Yes \\ 
        Controls & No & Yes & No & Yes\\
        \bottomrule
        \multicolumn{4}{l}{\emph{Clustered (Municipality) standard-errors in parentheses}}\\
   \multicolumn{4}{l}{\emph{Signif. Codes: ***: 0.01, **: 0.05, *: 0.1}}\\
        \end{tabular}\endgroup
        }
\end{table}

\subsection{Spillover effects}
In this section we investigate possible spillover effects of the dissolution of infiltrated councils to neighboring municipalities. In other words, we check whether the dissolution implied changes in the environmental policy in neighboring municipalities. There are good reasons to explore the existence of spillovers in the present context.

On the one hand, it is well known that, in general, the level of expenditures/taxes chosen by a municipality can affect the welfare of its neighbors \citep{gordon1983optimal}. In addition, when public expenditures present characteristics of complementarity, a decrease in spending in one municipality can lead to a decrease in its neighbors. The opposite can occur if the expenditure is for public goods that are substitutes \citep{case1993budget}. In the case of Italy, in particular, existing evidence suggests that municipalities can enjoy positive spillover effects from public expenditure of other municipalities: An increase in expenditures in one municipality can lead to an increase in public spending in its neighbors \citep{ferraresi2018spillover}. In the case of environmental policies, we saw in Section \ref{secEnvPolicies} that  decisions at different administrative levels are often interrelated, so that changes in policies in one municipality can trigger changes in the policy of other contiguous municipalities. 

Furthermore, in the specific context of waste management, interactions among local governments can exhibit characteristics of yardstick competition, where local performances are linked to the performance of their neighbors as politicians aim at maximizing the chances of re-election and their political consensus. In particular, evidence suggests that Italian provinces mimic their neighbors' behavior in terms of sorted waste collection, especially when the provincial President runs for re-election \citep{ferraresi2023waste}. Finally, in the context of the dissolution of infiltrated councils, some works have already detected the presence of spillovers. Specifically, \citet{galletta2017law} shows that the dissolution of a city council triggers a decrease in public investment in neighboring municipalities, while \citet{tulli2024sweeping} demonstrates that spillovers also occur in the implementation of policies on public contracts. All these results provide a justification for the analysis of spillovers

To identify neighboring municipalities, we considered the adjacency tables provided by ISTAT, indicating for each municipality the municipalities with neighboring administrative boundaries. With this information, we reestimated Eq. (\ref{eqATE}) assuming that a municipality is treated if one of its neighboring municipalities experienced a council dismissal. Tables \ref{tab:munexp_adj}  and \ref{tab:wman_adj} contain the results.

\begin{table}[H]
\caption{Municipal Expenditures (variables i) - iii)): treated municipalities are those with a dissolved council among the adjacent municipalities}
\label{tab:munexp_adj}
\centering
\adjustbox{max height=\dimexpr\textheight-10cm\relax, max width=14cm}{
\begingroup
%\renewcommand{\arraystretch}{1.2}
 % Prima sottotabella
        \begin{tabular}{lccccccccc}
        \toprule
        \multicolumn{5}{c}{\textit{ i) Capital Expenditures Sustainable Development of the Territory and the Environment}}\\
        \midrule
         &\multicolumn{2}{c}{\textit{Treatment starts after 3 years}} & \multicolumn{2}{c}{\textit{Treatment starts after dissolution}}\\
        \midrule\midrule
       Law164             & 0.5570$^{***}$ & 0.5186$^{***}$ & 0.5131$^{***}$ & 1.170$^{***}$  \\   
                      & (0.0764)       & (0.0756)       & (0.0695)       & (0.2318)  \\   
        \midrule
        \emph{Fit statistics} \\
        Observations       & 48,969         & 48,158         & 48,969         & 48,158\\  
        R$^2$     & 0.28063        & 0.29450        & 0.28034        & 0.29422\\  
   Within R$^2$   & 0.00213        & 0.01914        & 0.00173        & 0.01875\\  
        \midrule
         \multicolumn{5}{c}{\textit{ii) Current Expenditures Integrated Water Service}}\\
        \midrule
        &\multicolumn{2}{c}{\textit{Treatment starts after 3 years}} & \multicolumn{2}{c}{\textit{Treatment starts after dissolution}}\\
\midrule\midrule
        Law164        & 0.4899$^{***}$ & 0.4672$^{***}$ & 0.4328$^{***}$ & 0.4073$^{***}$  \\   
                      & (0.0598)       & (0.0599)       & (0.0584)       & (0.0589) \\   

        \midrule
        \emph{Fit statistics} \\
        Observations  & 79,674         & 78,801         & 79,674         & 78,801\\  
        R$^2$         & 0.70071        & 0.69434        & 0.70060        & 0.69422\\  
   Within R$^2$       & 0.00224        & 0.00372        & 0.00187        & 0.00332\\  
        \midrule
        \multicolumn{5}{c}{\textit{iii) Current Expenditures Green Urban Areas}}\\
        \midrule
        &\multicolumn{2}{c}{\textit{Treatment starts after 3 years}} & \multicolumn{2}{c}{\textit{Treatment starts after dissolution}}\\
   \midrule\midrule
   Law164             & 0.4043$^{***}$ & 0.3622$^{***}$ & 0.3738$^{***}$ & 0.3291$^{***}$  \\   
                      & (0.0702)       & (0.0703)       & (0.0657)       & (0.0663) \\   
   \midrule
   \emph{Fit statistics}\\
   Observations       & 79,675         & 78,802         & 79,675         & 78,802\\  
   R$^2$              & 0.59863        & 0.59823        & 0.59858        & 0.59817\\  
   Within R$^2$       & 0.00147        & 0.00419        & 0.00134        & 0.00404\\ 
   
        \midrule
        Municipality FE         & Yes & Yes & Yes & Yes \\  
        Year FE          & Yes & Yes & Yes & Yes \\ 
        Controls & No & Yes & No & Yes\\
       \bottomrule
        \multicolumn{4}{l}{\emph{Clustered (Municipality) standard-errors in parentheses}}\\
   \multicolumn{4}{l}{\emph{Signif. Codes: ***: 0.01, **: 0.05, *: 0.1}}\\
        \end{tabular}\endgroup
        }
\end{table}

The results in Table \ref{tab:munexp_adj} are consistent with those in Table \ref{tab:mun} with the addition that, in this case, the coefficient for the Current Expenditures on Green Areas is positive and significant.

The results in Table \ref{tab:wman_adj} are also broadly in line with the results in Table \ref{tab:wman}. The only remarkable difference appears in the estimation of the effect on the Costs of Recovery and Treatment of Sorted Waste (variable v)), which is positive and significant when the treatment is assumed to start right after the dissolution of the city council, and of the effect on the Total Cost of Waste Management, which always displays a non-significant effect.

\begin{table}[H]
\vspace{-3cm}
\caption{Waste Management Costs (Variables iv) - viii)): treated municipalities are those with a dissolved council in the adjacent municipalities
}
\label{tab:wman_adj}
\centering
\adjustbox{max height=\dimexpr\textheight-10cm\relax, max width=\textwidth}{
\begingroup
%\renewcommand{\arraystretch}{1.2}
 % Prima sottotabella
        \begin{tabular}{lcccc}
        \toprule
        \multicolumn{5}{c}{\textit{iv) Percentage of Sorted Waste}}\\
        \midrule
        &\multicolumn{2}{c}{\textit{Treatment starts after 3 years}} & \multicolumn{2}{c}{\textit{Treatment starts after dissolution}}\\
   \midrule \midrule
   Law164             & 0.0644$^{***}$ & 0.0592$^{***}$ & 0.0787$^{***}$ & 0.0587$^{***}$  \\   
                      & (0.0084)       & (0.0084)       & (0.0085)       & (0.0075)\\   
 
   \midrule

        \emph{Fit statistics}\\
   Observations       & 80,057         & 78,826         & 80,057         & 78,826\\  
   R$^2$              & 0.79798        & 0.79829        & 0.79846        & 0.79869\\  
   Within R$^2$       & 0.00400        & 0.00728        & 0.00634        & 0.00925\\  
  
   \midrule
         \multicolumn{5}{c}{\textit{v) Costs of Recovery and Treatment Sorted Waste}}\\
        \midrule
        &\multicolumn{2}{c}{\textit{Treatment starts after 3 years}} & \multicolumn{2}{c}{\textit{Treatment starts after dissolution}}\\
        \midrule\midrule
          Law164             & 0.0299                & 0.0124         & 0.1325$^{***}$ & 0.1174$^{**}$  \\   
                      & (0.0505)              & (0.0505)       & (0.0513)       & (0.0513)  \\   
   \midrule
\emph{Fit statistics}\\
   Observations       & 45,898                & 44,838         & 45,898         & 44,838\\  
   R$^2$              & 0.58275               & 0.58167        & 0.58293        & 0.58181\\  
   Within R$^2$       & $2.39\times 10^{-5}$  & 0.00332        & 0.00044        & 0.00367\\  
   \midrule
        \multicolumn{5}{c}{\textit{vi) Costs of Collection and Transport Unsorted Waste}}\\
        \midrule
        & \multicolumn{2}{c}{After Three Years} & \multicolumn{2}{c}{Full Lenght} \\
   \midrule\midrule

   Law164             & -0.0672$^{*}$ & -0.0679$^{*}$  & -0.0971$^{**}$ & -0.0970$^{**}$\\   
                      & (0.0362)      & (0.0363)       & (0.0420)       & (0.0420)\\ 
   \midrule
   \emph{Fit statistics}\\
  Observations        & 54,158        & 52,985         & 54,158         & 52,985\\  
   R$^2$              & 0.66653       & 0.66675        & 0.66658        & 0.66680\\  
   Within R$^2$       & 0.00015       & 0.00100        & 0.00030        & 0.00115\\  
\midrule
\multicolumn{5}{c}{\textit{vii) Costs of Collection and Transport Sorted Waste}}\\
\midrule
&\multicolumn{2}{c}{\textit{Treatment starts after 3 years}} & \multicolumn{2}{c}{\textit{Treatment starts after dissolution}}\\
   \midrule\midrule
   Law164             & 0.1552$^{***}$ & 0.1589$^{***}$ & 0.1789$^{***}$ & 0.1823$^{***}$\\   
                      & (0.0428)       & (0.0432)       & (0.0443)       & (0.0447)\\  
   \midrule
   \emph{Fit statistics}\\
   Observations       & 53,786         & 52,613         & 53,786         & 52,613\\  
   R$^2$              & 0.60263        & 0.59728        & 0.60270        & 0.59735\\  
   Within R$^2$       & 0.00067        & 0.00367        & 0.00086        & 0.00387\\ 
   \midrule
   \multicolumn{5}{c}{\textit{viii) Total Costs of Waste Management}}\\
   \midrule
   &\multicolumn{2}{c}{\textit{Treatment starts after 3 years}} & \multicolumn{2}{c}{\textit{Treatment starts after dissolution}}\\
   \midrule\midrule
   \emph{Variables}\\
   Law164             & -0.0010               & 0.0020        & 0.0046                & 0.0055\\   
                      & (0.0120)              & (0.0120)      & (0.0145)              & (0.0146)\\   
  \midrule
   \emph{Fit statistics}\\
    Observations       & 55,744                & 54,559        & 55,744                & 54,559\\  
   R$^2$              & 0.85055               & 0.84735       & 0.85055               & 0.84735\\  
   Within R$^2$       & $3.18\times 10^{-7}$  & 0.00346       & $6.12\times 10^{-6}$  & 0.00347\\
   \midrule
         Municipality FE         & Yes & Yes & Yes & Yes \\  
        Year FE          & Yes & Yes & Yes & Yes \\ 
        Controls & No & Yes & No & Yes\\
        \bottomrule
        \multicolumn{4}{l}{\emph{Clustered (Municipality) standard-errors in parentheses}}\\
   \multicolumn{4}{l}{\emph{Signif. Codes: ***: 0.01, **: 0.05, *: 0.1}}\\
        \end{tabular}\endgroup
        }
\end{table}

\subsection{Discussion}
In general, our results suggest that the removal of Mafia-infiltrated city councils improves the environmental performance of local administrations, particularly in key areas such as sustainable waste management. In some cases, in particular, the event study highlights that some time after the dissolution is needed for the effect to become visible. This is the case of Capital Expenditures on Sustainable Development and the Environment (variable i)) and of the Percentage of Sorted Waste (variable iv)).

With respect to the costs associated to waste management, we highlight a clear change in the expenditures after the dissolution and, eventually, after a new city council is elected, with a clear change of priority towards sorted waste at the expenses of unsorted wastes.

These findings confirm our initial conjecture. When a city council is infiltrated by organized crime, its environmental policies are not efficient. After dissolution, the rise in treatment and recovery costs likely reflects a reallocation of resources toward more sustainable waste management practices. We interpret these results as confirming that criminal organizations do not have an interest in efficient environmental policies as such policies do not increase their profits as inefficient policies. In fact, illegal dumping and trafficking of waste are known lines of criminal business for Mafia organizations, which can therefore thrive when environmental policies are inefficient.

The environmental policies implemented after the dissolution clearly imply an improvement in the environmental quality but, as existing evidence suggest, they can also imply an overall better management of the environment-related activities. In fact, sorted waste collection is generally more expensive and offers fewer economies of scale than unsorted waste \citep{greco2015drivers,d2016full,bartolacci2019efficiency}. However, increasing the proportion of separated waste can reduce collection costs over time and result in general cost advantages \citep{d2016full,bartolacci2019efficiency,cialani2020cost}. Effective waste separation and collection are key to achieving efficient waste management \citep{lo2021effectiveness}, and their implementation can also improve the financial performance of waste management companies \citep{bartolacci2018assessing,bartolacci2018relationship}.

Adopting more circular treatment options has been shown to be cost-effective for overall waste management systems \citep{beigl2004comparison,lavee2007municipal,di2020drivers}. For example, directing unsorted waste to waste-to-energy plants instead of landfills can lower overall waste management costs \citep{beigl2004comparison,cucchiella2014strategic,di2020drivers}. Integrated waste management systems generate the greatest cost savings and environmental benefits \citep{emery2007environmental,abrate2014costs}. The joint offering of disposal and recycling services enables the creation of scope economies, resulting in true cost savings for municipalities \citep{callan2001economies}. Large-capacity recovery plants for sorted waste tend to be more efficient than traditional “collect-and-landfill” systems \citep{lombrano2009cost}. This may also explain the observed behavior of unsorted waste collection and transport costs (see Figure \ref{fig:unsortcosts}): these costs begin to rise two years after dissolution but then decline around the same time as those for sorted waste. The presence and size of consortia also influences costs; in the Italian context, integrated collection is cost-effective only for small consortia \citep{guerrini2017assessing}.

Regarding municipal expenditures, the only variable that shows no significant effect post-dissolution is spending on green urban areas. This may be due to institutional reforms that shifted responsibilities for green space management to newly established sub-regional administrative levels, such as the \textit{Citta' Metropolitane} (Metropolitan Cities). These entities hold broader competencies and access to funding for the management of urban and peri-urban green areas. The lack of a significant effect at the municipal level may reflect the increased role and fiscal capacity of Metropolitan Cities in this domain.\footnote{For some areas of expenditures the effect of infiltration at the municipal level was reduced due to institutional changes. After the introduction of Legislative Decree 152/2006, the competencies of municipalities in areas such as water management and green urban areas have been progressively transferred to other administrative levels (e.g., Metropolitan Cities, EGATO)} 

Our results also show that dissolution of municipal councils due to mafia infiltration improves environmental performance in neighboring municipalities. Thus, neighboring municipalities benefit from the positive effects of improving the environmental performance of dissolved municipalities. In particular, the results on expenditures for urban green areas reveal an interesting dynamic. Although municipalities directly affected by a dissolution due to mafia infiltration do not show a significant change in green area spending, the analysis of spillover effects indicates that neighboring municipalities exhibit a positive and significant increase in such expenditures. This pattern may reflect a strategic response: municipalities bordering a dissolved entity might invest more in urban green spaces as a visible signal of good governance, a preventive measure to enhance social cohesion and civic presence, and a way to distance themselves reputationally from corruption-related risks. Such spending may also serve to reinforce public trust and reduce the informal appropriation of public space in areas where institutional credibility is under scrutiny. It might be also related to a different distribution of competencies between municipalities and higher levels of administration.

\section{Conclusions\label{secConclusions}}
This work aims to assess the role of mafia infiltration in the environmental performance of municipalities. Organized crime is a pervasive phenomenon in some Italian regions and in other parts of the world that affects all aspects of the socioeconomic fabric. A relevant amount of work assessed the effect of organized crime infiltration on the socioeconomic outcomes of a territory. However, few studies have investigated the organized crime role in endangering local sustainable development. Our results, therefore, suggest another negative effect of criminal organizations, namely, a lower capacity of local governments to efficiently and effectively deal with environmental issues, with straightforward policy implications. 

Municipalities have competencies on environmental management directly related to providing services to citizens, such as waste management. In general, the municipalities dissolved for mafia infiltration experience an increase in environmental-related expenditures. Thus, organized crime keeps environmental management ineffective and inefficient. This work suggests another reason why the contrast of criminal organizations is of paramount importance for socioeconomic development.

\bibliography{references.bib}

\begin{thebibliography}{}

\bibitem[\protect\citeauthoryear{Abrate, Erbetta, Fraquelli, and Vannoni}{Abrate et~al.}{2014}]{abrate2014costs}
Abrate, G., F.~Erbetta, G.~Fraquelli, and D.~Vannoni (2014).
\newblock The costs of disposal and recycling: an application to italian municipal solid waste services.
\newblock {\em Regional Studies\/}~{\em 48\/}(5), 896--909.

\bibitem[\protect\citeauthoryear{Acconcia, Corsetti, and Simonelli}{Acconcia et~al.}{2014}]{acconcia2014mafia}
Acconcia, A., G.~Corsetti, and S.~Simonelli (2014).
\newblock Mafia and public spending: Evidence on the fiscal multiplier from a quasi-experiment.
\newblock {\em American Economic Review\/}~{\em 104\/}(7), 2185--2209.

\bibitem[\protect\citeauthoryear{Alesina, Piccolo, and Pinotti}{Alesina et~al.}{2019}]{alesina2019organized}
Alesina, A., S.~Piccolo, and P.~Pinotti (2019).
\newblock Organized crime, violence, and politics.
\newblock {\em The Review of Economic Studies\/}~{\em 86\/}(2), 457--499.

\bibitem[\protect\citeauthoryear{Ashraf and Weil}{Ashraf and Weil}{2024}]{ashraf2024economic}
Ashraf, Q.~H. and D.~N. Weil (2024).
\newblock {\em Economic growth}.
\newblock Taylor \& Francis.

\bibitem[\protect\citeauthoryear{Baraldi, Cantabene, and De~Iudicibus}{Baraldi et~al.}{2024}]{baraldi2024fighting}
Baraldi, A.~L., C.~Cantabene, and A.~De~Iudicibus (2024).
\newblock Fighting crime to improve recycling: Evaluating an anti-mafia policy on source separation of waste.
\newblock {\em Ecological Economics\/}~{\em 224}, 108291.

\bibitem[\protect\citeauthoryear{Baraldi, Immordino, and Stimolo}{Baraldi et~al.}{2022}]{baraldi2022self}
Baraldi, A.~L., G.~Immordino, and M.~Stimolo (2022).
\newblock Self-selecting candidates or compelling voters: How organized crime affects political selection.
\newblock {\em European Journal of Political Economy\/}~{\em 71}, 102133.

\bibitem[\protect\citeauthoryear{Bartolacci, Del~Gobbo, Paolini, and Soverchia}{Bartolacci et~al.}{2019}]{bartolacci2019efficiency}
Bartolacci, F., R.~Del~Gobbo, A.~Paolini, and M.~Soverchia (2019).
\newblock Efficiency in waste management companies: A proposal to assess scale economies.
\newblock {\em Resources, Conservation and Recycling\/}~{\em 148}, 124--131.

\bibitem[\protect\citeauthoryear{Bartolacci, Paolini, Quaranta, and Soverchia}{Bartolacci et~al.}{2018a}]{bartolacci2018assessing}
Bartolacci, F., A.~Paolini, A.~G. Quaranta, and M.~Soverchia (2018a).
\newblock Assessing factors that influence waste management financial sustainability.
\newblock {\em Waste management\/}~{\em 79}, 571--579.

\bibitem[\protect\citeauthoryear{Bartolacci, Paolini, Quaranta, and Soverchia}{Bartolacci et~al.}{2018b}]{bartolacci2018relationship}
Bartolacci, F., A.~Paolini, A.~G. Quaranta, and M.~Soverchia (2018b).
\newblock The relationship between good environmental practices and financial performance: Evidence from italian waste management companies.
\newblock {\em Sustainable Production and Consumption\/}~{\em 14}, 129--135.

\bibitem[\protect\citeauthoryear{Beigl and Salhofer}{Beigl and Salhofer}{2004}]{beigl2004comparison}
Beigl, P. and S.~Salhofer (2004).
\newblock Comparison of ecological effects and costs of communal waste management systems.
\newblock {\em Resources, Conservation and Recycling\/}~{\em 41\/}(2), 83--102.

\bibitem[\protect\citeauthoryear{Borusyak, Jaravel, and Spiess}{Borusyak et~al.}{2024}]{borusyak2024revisiting}
Borusyak, K., X.~Jaravel, and J.~Spiess (2024).
\newblock Revisiting event-study designs: robust and efficient estimation.
\newblock {\em Review of Economic Studies\/}, rdae007.

\bibitem[\protect\citeauthoryear{Brombacher, Garz{\'o}n, and V{\'e}lez}{Brombacher et~al.}{2021}]{brombacher2021introduction}
Brombacher, D., J.~C. Garz{\'o}n, and M.~A. V{\'e}lez (2021).
\newblock Introduction special issue: Environmental impacts of illicit economies.
\newblock {\em Journal of Illicit Economies and Development\/}~{\em 3\/}(1).

\bibitem[\protect\citeauthoryear{Buonanno, Ferrari, and Saia}{Buonanno et~al.}{2024}]{buonanno2024all}
Buonanno, P., I.~Ferrari, and A.~Saia (2024).
\newblock All is not lost: Organized crime and social capital formation.
\newblock {\em Journal of Public Economics\/}~{\em 240}, 105257.

\bibitem[\protect\citeauthoryear{Buonanno and Pazzona}{Buonanno and Pazzona}{2014}]{buonannoPazzona2014}
Buonanno, P. and M.~Pazzona (2014).
\newblock Migrating mafias.
\newblock {\em Regional Science and Urban Economics\/}~{\em 44}, 75--81.

\bibitem[\protect\citeauthoryear{Callan and Thomas}{Callan and Thomas}{2001}]{callan2001economies}
Callan, S.~J. and J.~M. Thomas (2001).
\newblock Economies of scale and scope: A cost analysis of municipal solid waste services.
\newblock {\em Land economics\/}~{\em 77\/}(4), 548--560.

\bibitem[\protect\citeauthoryear{Callaway and Sant'Anna}{Callaway and Sant'Anna}{2021}]{CallawaySantAnna2021}
Callaway, B. and P.~H. Sant'Anna (2021).
\newblock Difference-in-differences with multiple time periods.
\newblock {\em Journal of Econometrics\/}~{\em 225\/}(2), 200--230.

\bibitem[\protect\citeauthoryear{Case, Rosen, and Hines~Jr}{Case et~al.}{1993}]{case1993budget}
Case, A.~C., H.~S. Rosen, and J.~R. Hines~Jr (1993).
\newblock Budget spillovers and fiscal policy interdependence: Evidence from the states.
\newblock {\em Journal of public economics\/}~{\em 52\/}(3), 285--307.

\bibitem[\protect\citeauthoryear{Cialani and Mortazavi}{Cialani and Mortazavi}{2020}]{cialani2020cost}
Cialani, C. and R.~Mortazavi (2020).
\newblock The cost of urban waste management: An empirical analysis of recycling patterns in italy.
\newblock {\em Frontiers in Sustainable Cities\/}~{\em 2}, 8.

\bibitem[\protect\citeauthoryear{Cingano and Tonello}{Cingano and Tonello}{2020}]{cingano2020law}
Cingano, F. and M.~Tonello (2020).
\newblock Law enforcement, social control and organized crime: Evidence from local government dismissals in italy.
\newblock {\em Italian Economic Journal\/}~{\em 6\/}(2), 221--254.

\bibitem[\protect\citeauthoryear{Cucchiella, D’Adamo, and Gastaldi}{Cucchiella et~al.}{2014}]{cucchiella2014strategic}
Cucchiella, F., I.~D’Adamo, and M.~Gastaldi (2014).
\newblock Strategic municipal solid waste management: A quantitative model for italian regions.
\newblock {\em Energy Conversion and Management\/}~{\em 77}, 709--720.

\bibitem[\protect\citeauthoryear{Daniele and Geys}{Daniele and Geys}{2015}]{daniele2015organised}
Daniele, G. and B.~Geys (2015).
\newblock Organised crime, institutions and political quality: Empirical evidence from italian municipalities.
\newblock {\em The Economic Journal\/}~{\em 125\/}(586), F233--F255.

\bibitem[\protect\citeauthoryear{Devine, Wrathall, Aguilar-Gonz{\'a}lez, Benessaiah, Tellman, Ghaffari, and Ponstingel}{Devine et~al.}{2021}]{devine2021narco}
Devine, J.~A., D.~Wrathall, B.~Aguilar-Gonz{\'a}lez, K.~Benessaiah, B.~Tellman, Z.~Ghaffari, and D.~Ponstingel (2021).
\newblock Narco-degradation: Cocaine trafficking’s environmental impacts in central america’s protected areas.
\newblock {\em World Development\/}~{\em 144}, 105474.

\bibitem[\protect\citeauthoryear{Di~Cataldo and Mastrorocco}{Di~Cataldo and Mastrorocco}{2022}]{di2022organized}
Di~Cataldo, M. and N.~Mastrorocco (2022).
\newblock Organized crime, captured politicians, and the allocation of public resources.
\newblock {\em The Journal of Law, Economics, and Organization\/}~{\em 38\/}(3), 774--839.

\bibitem[\protect\citeauthoryear{Di~Foggia and Beccarello}{Di~Foggia and Beccarello}{2020}]{di2020drivers}
Di~Foggia, G. and M.~Beccarello (2020).
\newblock Drivers of municipal solid waste management cost based on cost models inherent to sorted and unsorted waste.
\newblock {\em Waste Management\/}~{\em 114}, 202--214.

\bibitem[\protect\citeauthoryear{Di~Pillo, Levialdi, and Marzano}{Di~Pillo et~al.}{2023}]{di2023organized}
Di~Pillo, F., N.~Levialdi, and R.~Marzano (2023).
\newblock Organized crime and waste management costs.
\newblock {\em Regional Studies\/}~{\em 57\/}(1), 168--180.

\bibitem[\protect\citeauthoryear{D'Onza, Greco, and Allegrini}{D'Onza et~al.}{2016}]{d2016full}
D'Onza, G., G.~Greco, and M.~Allegrini (2016).
\newblock Full cost accounting in the analysis of separated waste collection efficiency: A methodological proposal.
\newblock {\em Journal of environmental management\/}~{\em 167}, 59--65.

\bibitem[\protect\citeauthoryear{D’Amato, Mazzanti, and Nicolli}{D’Amato et~al.}{2015}]{d2015waste}
D’Amato, A., M.~Mazzanti, and F.~Nicolli (2015).
\newblock Waste and organized crime in regional environments: How waste tariffs and the mafia affect waste management and disposal.
\newblock {\em Resource and energy economics\/}~{\em 41}, 185--201.

\bibitem[\protect\citeauthoryear{Emery, Davies, Griffiths, and Williams}{Emery et~al.}{2007}]{emery2007environmental}
Emery, A., A.~Davies, A.~Griffiths, and K.~Williams (2007).
\newblock Environmental and economic modelling: A case study of municipal solid waste management scenarios in wales.
\newblock {\em Resources, conservation and recycling\/}~{\em 49\/}(3), 244--263.

\bibitem[\protect\citeauthoryear{Fenizia and Saggio}{Fenizia and Saggio}{2024}]{fenizia2024organized}
Fenizia, A. and R.~Saggio (2024).
\newblock Organized crime and economic growth: evidence from municipalities infiltrated by the mafia.
\newblock Technical report, National Bureau of Economic Research.

\bibitem[\protect\citeauthoryear{Ferraresi, Mazzanti, Mazzarano, Rizzo, and Secomandi}{Ferraresi et~al.}{2023}]{ferraresi2023waste}
Ferraresi, M., M.~Mazzanti, M.~Mazzarano, L.~Rizzo, and R.~Secomandi (2023).
\newblock Waste recycling and yardstick competition among italian provinces after the eu waste framework directive.
\newblock {\em Regional Studies\/}~{\em 57\/}(8), 1535--1545.

\bibitem[\protect\citeauthoryear{Ferraresi, Migali, and Rizzo}{Ferraresi et~al.}{2018}]{ferraresi2018spillover}
Ferraresi, M., G.~Migali, and L.~Rizzo (2018).
\newblock Spillover effects in local public spending.
\newblock {\em Regional Studies\/}~{\em 52\/}(11), 1570--1584.

\bibitem[\protect\citeauthoryear{Galletta}{Galletta}{2017}]{galletta2017law}
Galletta, S. (2017).
\newblock Law enforcement, municipal budgets and spillover effects: Evidence from a quasi-experiment in italy.
\newblock {\em Journal of Urban Economics\/}~{\em 101}, 90--105.

\bibitem[\protect\citeauthoryear{Germani, Pergolizzi, and Reganati}{Germani et~al.}{2018}]{germani2018eco}
Germani, A.~R., A.~Pergolizzi, and F.~Reganati (2018).
\newblock Eco-mafia and environmental crime in italy.
\newblock {\em Green crimes and dirty money\/}, 42--71.

\bibitem[\protect\citeauthoryear{Goodman-Bacon}{Goodman-Bacon}{2021}]{GoodmanBacon2021}
Goodman-Bacon, A. (2021).
\newblock Difference-in-differences with variation in treatment timing.
\newblock {\em Journal of Econometrics\/}~{\em 225\/}(2), 254--277.

\bibitem[\protect\citeauthoryear{Gordon}{Gordon}{1983}]{gordon1983optimal}
Gordon, R.~H. (1983).
\newblock An optimal taxation approach to fiscal federalism.
\newblock {\em The Quarterly Journal of Economics\/}~{\em 98\/}(4), 567--586.

\bibitem[\protect\citeauthoryear{Greco, Allegrini, Del~Lungo, Savellini, and Gabellini}{Greco et~al.}{2015}]{greco2015drivers}
Greco, G., M.~Allegrini, C.~Del~Lungo, P.~G. Savellini, and L.~Gabellini (2015).
\newblock Drivers of solid waste collection costs. empirical evidence from italy.
\newblock {\em Journal of Cleaner Production\/}~{\em 106}, 364--371.

\bibitem[\protect\citeauthoryear{Guerrini, Carvalho, Romano, Marques, and Leardini}{Guerrini et~al.}{2017}]{guerrini2017assessing}
Guerrini, A., P.~Carvalho, G.~Romano, R.~C. Marques, and C.~Leardini (2017).
\newblock Assessing efficiency drivers in municipal solid waste collection services through a non-parametric method.
\newblock {\em Journal of cleaner production\/}~{\em 147}, 431--441.

\bibitem[\protect\citeauthoryear{Hork{\`y}, Grabic, Grabicov{\'a}, Brooks, Douda, Slav{\'\i}k, Huben{\'a}, Sancho~Santos, and Rand{\'a}k}{Hork{\`y} et~al.}{2021}]{horky2021methamphetamine}
Hork{\`y}, P., R.~Grabic, K.~Grabicov{\'a}, B.~W. Brooks, K.~Douda, O.~Slav{\'\i}k, P.~Huben{\'a}, E.~M. Sancho~Santos, and T.~Rand{\'a}k (2021).
\newblock Methamphetamine pollution elicits addiction in wild fish.
\newblock {\em Journal of Experimental Biology\/}~{\em 224\/}(13), jeb242145.

\bibitem[\protect\citeauthoryear{Kangaspunta and Marshall}{Kangaspunta and Marshall}{2009}]{kangaspunta2009eco}
Kangaspunta, K. and I.~H. Marshall (2009).
\newblock Eco-crime and justice.
\newblock {\em Essays on environmental crime. Turin: UNICRI\/}.

\bibitem[\protect\citeauthoryear{Kennedy}{Kennedy}{2009}]{BBC2009}
Kennedy, D. (2009).
\newblock Mafia 'sank ships of toxic waste'.

\bibitem[\protect\citeauthoryear{Lavee}{Lavee}{2007}]{lavee2007municipal}
Lavee, D. (2007).
\newblock Is municipal solid waste recycling economically efficient?
\newblock {\em Environmental Management\/}~{\em 40}, 926--943.

\bibitem[\protect\citeauthoryear{Lavezzi}{Lavezzi}{2014}]{lavezzi2014organised}
Lavezzi, A.~M. (2014).
\newblock Organised crime and the economy: A framework for policy prescriptions.
\newblock {\em Global Crime\/}~{\em 15\/}(1-2), 164--190.

\bibitem[\protect\citeauthoryear{Legambiente}{Legambiente}{2024}]{legambiente2024ecomafia}
Legambiente (2024).
\newblock {\em Ecomafia 2024}.
\newblock Edizioni Ambiente.

\bibitem[\protect\citeauthoryear{lo~Storto}{lo~Storto}{2021}]{lo2021effectiveness}
lo~Storto, C. (2021).
\newblock Effectiveness-efficiency nexus in municipal solid waste management: A non-parametric evidence-based study.
\newblock {\em Ecological Indicators\/}~{\em 131}, 108185.

\bibitem[\protect\citeauthoryear{Lombrano}{Lombrano}{2009}]{lombrano2009cost}
Lombrano, A. (2009).
\newblock Cost efficiency in the management of solid urban waste.
\newblock {\em Resources, Conservation and Recycling\/}~{\em 53\/}(11), 601--611.

\bibitem[\protect\citeauthoryear{Lynch}{Lynch}{2020}]{lynch2020green}
Lynch, M.~J. (2020).
\newblock Green criminology and environmental crime: Criminology that matters in the age of global ecological collapse.
\newblock {\em Journal of White Collar and Corporate Crime\/}~{\em 1\/}(1), 50--61.

\bibitem[\protect\citeauthoryear{Massari and Monzini}{Massari and Monzini}{2004}]{massari2004dirty}
Massari, M. and P.~Monzini (2004).
\newblock Dirty businesses in italy: A case-study of illegal trafficking in hazardous waste.
\newblock {\em Global crime\/}~{\em 6\/}(3-4), 285--304.

\bibitem[\protect\citeauthoryear{Pardal, Colman, and Surmont}{Pardal et~al.}{2021}]{pardal2021synthetic}
Pardal, M., C.~Colman, and T.~Surmont (2021).
\newblock Synthetic drug production in belgium: environmental harms as collateral damage?
\newblock {\em Journal of Illicit Economies and Development\/}~{\em 3\/}(1), 36--49.

\bibitem[\protect\citeauthoryear{Pinotti}{Pinotti}{2015a}]{pinotti2015causes}
Pinotti, P. (2015a).
\newblock The causes and consequences of organised crime: Preliminary evidence across countries.
\newblock {\em The Economic Journal\/}~{\em 125\/}(586), F158--F174.

\bibitem[\protect\citeauthoryear{Pinotti}{Pinotti}{2015b}]{pinotti2015economic}
Pinotti, P. (2015b).
\newblock The economic costs of organised crime: Evidence from southern italy.
\newblock {\em The Economic Journal\/}~{\em 125\/}(586), F203--F232.

\bibitem[\protect\citeauthoryear{Ravenda, Giuranno, Valencia-Silva, Argiles-Bosch, and Garc{\'\i}a-Bland{\'o}n}{Ravenda et~al.}{2020}]{ravenda2020effects}
Ravenda, D., M.~G. Giuranno, M.~M. Valencia-Silva, J.~M. Argiles-Bosch, and J.~Garc{\'\i}a-Bland{\'o}n (2020).
\newblock The effects of mafia infiltration on public procurement performance.
\newblock {\em European Journal of Political Economy\/}~{\em 64}, 101923.

\bibitem[\protect\citeauthoryear{Rege and Lavorgna}{Rege and Lavorgna}{2017}]{rege2017organization}
Rege, A. and A.~Lavorgna (2017).
\newblock Organization, operations, and success of environmental organized crime in italy and india: A comparative analysis.
\newblock {\em European Journal of Criminology\/}~{\em 14\/}(2), 160--182.

\bibitem[\protect\citeauthoryear{Reitano}{Reitano}{2018}]{reitano2018organized}
Reitano, T. (2018).
\newblock Organized crime as a threat to sustainable development: Understanding the evidence.
\newblock {\em Organized Crime and Illicit Trade: How to Respond to This Strategic Challenge in Old and New Domains\/}, 23--35.

\bibitem[\protect\citeauthoryear{Saviano}{Saviano}{2006}]{saviano2006gomorra}
Saviano, R. (2006).
\newblock {\em Gomorra: A Personal Journey into the Violent International Empire of Naples' Organized Crime System}.
\newblock Mondadori.

\bibitem[\protect\citeauthoryear{Sergi and South}{Sergi and South}{2016}]{sergi2016earth}
Sergi, A. and N.~South (2016).
\newblock ‘earth, water, air, and fire’: Environmental crimes, mafia power and political negligence in calabria.
\newblock {\em Illegal entrepreneurship, organized crime and social control: Essays in honor of professor Dick Hobbs\/}, 85--100.

\bibitem[\protect\citeauthoryear{Thompson}{Thompson}{2023}]{thompson2023convergence}
Thompson, S.~T. (2023).
\newblock The convergence of environmental crime and corruption: An operational typology.
\newblock {\em International Criminology\/}~{\em 3\/}(2), 133--148.

\bibitem[\protect\citeauthoryear{Tulli}{Tulli}{2024}]{tulli2024sweeping}
Tulli, A. (2024).
\newblock Sweeping the dirt under the rug: measuring spillovers of an anti-corruption measure.
\newblock {\em The Journal of Law, Economics, and Organization\/}, ewae009.

\bibitem[\protect\citeauthoryear{Unep and ASSESSMENT}{Unep and ASSESSMENT}{2016}]{unep2016rise}
Unep, A. and I.~R.~R. ASSESSMENT (2016).
\newblock The rise of environmental crime.
\newblock {\em Nairobi: UNEP\/}.

\bibitem[\protect\citeauthoryear{Varese}{Varese}{2011}]{varese2011mafias}
Varese, F. (2011).
\newblock {\em Mafias on the move: How organized crime conquers new territories}.
\newblock Princeton University Press.

\bibitem[\protect\citeauthoryear{Zabyelina and van Uhm}{Zabyelina and van Uhm}{2020}]{zabyelina2020new}
Zabyelina, Y. and D.~van Uhm (2020).
\newblock The new eldorado: Organized crime, informal mining, and the global scarcity of metals and minerals.
\newblock {\em Illegal mining: Organized crime, corruption, and ecocide in a resource-scarce world\/}, 3--30.

\end{thebibliography}

\end{document}